\documentclass[twocolumn,a4paper,10pt]{article}
\pdfoutput=1
\usepackage{fullpage}
\usepackage{cite}
\usepackage{times}

\usepackage{graphicx} 

\graphicspath{{images/}}

\usepackage{times}                 
\usepackage[british]{babel}        
\usepackage{graphicx}              %
\usepackage{amsmath}               
\usepackage{url}                   
\usepackage[footnotesize, bf]{caption}
\usepackage[font=footnotesize]{subfig} 
\usepackage{acronym}               
\usepackage{booktabs}              
\usepackage{bytefield}             
\usepackage{listings}              
\usepackage{algorithm}             
\usepackage{algorithmic}           

\begin{document}

\title{Characterizing Internet Video for Large-scale Active Measurements}

\author{ Saba Ahsan \\Aalto University     \\saba.ahsan@aalto.fi
  \and Varun Singh \\Aalto University     \\varun.singh@aalto.fi
  \and J\"org Ott   \\Aalto University     \\jorg.ott@aalto.fi
}
\date{5 August 2014}
%

\maketitle
\begin{abstract}


The availability of high definition video content on the web has
brought about a significant change in the characteristics of Internet
video, but not many studies on characterizing video have been done
after this change. Video characteristics such as video length, format,
target bit rate, and resolution provide valuable input to design
Adaptive Bit Rate (ABR) algorithms, sizing playout buffers in Dynamic
Adaptive HTTP streaming (DASH) players, model the variability in video
frame sizes, etc.  This paper presents datasets collected in 2013 and
2014 that contains over 130,000 videos from YouTube's most viewed (or
most popular) video charts in 58 countries.  We describe the basic
characteristics of the videos on YouTube for each category, format,
video length, file size, and data rate variation, observing that video
length and file size fit a log normal distribution.  We show that
three minutes of a video suffice to represent its instant data rate
fluctuation and that we can infer data rate characteristics of
different video resolutions from a single given one.  Based on our
findings, we design active measurements for measuring the performance
of Internet video.

\end{abstract}

\section{Introduction}

ISPs and application service providers have a strong interest in understanding
network and application performance to make sure that their customers are
satisfied.  In addition to passive traffic monitoring inside the network,
performing active measurements at the endpoints is gaining importance
as a tool for observing long-term network behavior as well as for
investigating and diagnosing network failures.  Measurement endpoints
include infrastructure nodes such as access routers and set-top boxes
as well as user devices such as personal computers, smartphones, and
tablets.  Typical metrics, e.g., as defined by the IP Performance
Metrics (IPPM) Working
Group\footnote{https://datatracker.ietf.org/wg/ippm/} are round trip
delay, one way delay, IP packet delay variation, average TCP/UDP
throughput, average fractional loss, DNS latency, among others.
Aggregating performance metrics from many measurement points by an
Internet Service Provider (ISP), or a measurement service (e.g., RIPE
Atlas\footnote{https://atlas.ripe.net/},
SamKnows\footnote{http://www.samknows.com/broadband/},
Netradar\footnote{https://www.netradar.org/}~\cite{netradar},
Speedtest\footnote{http://www.speedtest.com/}, etc.) allows
characterizing the network performance geo-spatially and over time,
diagnose outages and observe the impact of the outage, and lastly the
collected information helps regulators develop better public policy
for the Internet.


Currently, video is the dominant traffic on the Internet, in both fixed and
wireless networks. In 2012, 51\,\% of mobile traffic was
video~\cite{index520862global,dawn-zb}, hence, measuring the performance of
video streaming applications is crucial for ISPs. The video quality at an
endpoint is affected by path capacity (e.g., media bit rate is higher
than the available end-to-end capacity), burstiness of video (e.g., high
motion in the video causes temporary increase in media bit rate, which 
appears as a traffic burst on the network), network
packet loss and re-ordering. Therefore, transport layer metrics provide valuable
input to measuring a viewer's users experience~\cite{draft.ippm.stream}.
Performing large-scale passive measurements raises privacy concerns, because
end-users do not want the ISP or the measurement service to monitor their
traffic. Furthermore, metrics from a passive measurement are hard to correlate
across measurement points because there might be varying amounts of 
cross-traffic, which would be difficult to reconcile during analysis.

The motivation for collecting the datasets presented in this paper is to
explore the characteristics of Internet video for the design of active
measurement techniques at the endpoint, which is
suitable for large scale measurements (as defined by the IETF
LMAP\footnote{https://datatracker.ietf.org/wg/lmap/} Working Group). In order
to reflect user experience, the measurements are based on actual online videos
that are popular amongst users, instead of using a single predefined video. However,
the diversity in the duration, types and formats of videos available on the
Internet makes it hard to select an appropriate video for benchmarking user's
quality of experience. Furthermore, results from tests conducted on different
videos cannot be compared directly with each other. In this paper, we
present the analysis of datasets of YouTube's popular videos collected
between July 2013 to April 2014. We choose YouTube for two
reasons:  ease of access without logging in and its popularity\cite{sandvine2012global}. The 
datasets\footnote{http://www.netlab.tkk.fi/tutkimus/rtc/} contain
information collected for over \emph{130000} videos from 58 locations using YouTube's location-based charts. This paper makes the following four contributions:

\begin{enumerate}
\itemsep0pt \parskip0pt \parsep0pt
  \item We describe the video trends in terms of categories, duration,
formats, resolutions, media bit rates and the variation in instantaneous bit rates of the video (burstiness) in the current Internet. These results can be used for selecting appropriate videos for conducting active measurements. 
  \item We show that video lengths on YouTube follow a lognormal distribution. Additionally, the file sizes of different file formats (WebM and MP4) and resolutions (360p, 720p and 1080p) also follow lognormal distribution.  
  \item We observe that the average bit rate and the burstiness of a video when calculated for the first 3 minutes is comparable to the entire duration of the video (typically at least 10 minutes long). Since, the time taken and the traffic generated by active tests need to be minimized to avoid any effect on real user traffic, we can use 3 minutes as the cut-off time for our measurements. 
  \item  We show correlation of videos across different resolutions, arguing that it is possible to generate traffic for a higher resolution stream from a lower one or vice versa, by appropriate upscaling or downscaling, respectively. 
\end{enumerate}

The rest of this paper is organized as follows. We describe related work and the novelty of our work in 
Section~\ref{sec_relatedwork}. Section~\ref{dataset} describes our datasets and the methodology used for the collection. Results and analysis are divided into Section ~\ref{analysis1} and ~\ref{analysis2} followed by a discussion on the application of our results and future work in Section ~\ref{discussion}. We present a model for active measurements that is based on our current setup in Section \ref{lmapmodel} and conclude with a brief summary and hints at future work in Section~\ref{conclusion}.

\section{Related Work}
\label{sec_relatedwork}

There are several other studies that characterize YouTube videos but the datasets are from 2007-08~\cite{gill2007youtube, adhikari2010youtube, cheng2013understanding}. Initially in 2007, YouTube had a size limit of 100MB~\cite{gill2007youtube} for its videos, which has since been increased to 20GB~\cite{youtube-limit}. Our datasets were collected in 2013 and 2014 and contains Full HD (1080p) content as well as files with the WebM format, which to our knowledge has not been studied before.  

In \cite{brodersen2012youtube}, the authors use over 20 million randomly selected YouTube videos to show that the popularity of videos is constrained by geographical locations. Our methodology is in line with this as we gathered all available location-based charts from YouTube, giving our dataset regional representation. Furthermore, our proposal to LMAP for testing video streaming also recommends using location-based charts for measuring user experience. 

A crowdsourcing study in \cite{hossfeld2011quantification} shows that the QoE for TCP video streaming is directly related to the number and duration of stalls during a video playout. In \cite{casas2013monitoring}, the authors build a QoE model based on stalling events for YouTube. Research has also shown that actively measuring stall events (with the Pytomo tool~\cite{juluri2013viewing}) in different ISPs helps predicting the user experience~\cite{plissonneau2012analyzing}. The proposals in this paper can complement such a tool (like Pytomo) by selecting and categorizing videos for active measurements. 

A more recent study was done for the characterization of an adult video streaming website \cite{tyson2013demystifying}: the authors' findings about the video durations is similar to what we observe in our dataset; however, we offer an additional in-depth analysis of formats, resolutions and variations in the instantaneous bit rate. Since YouTube dominates video traffic, our findings can serve as a good comparison point for similar studies on other video streaming services.

In \cite{alcock2011application}, the researchers study how YouTube's block-sending flow control can lead to TCP packet losses. The impact of location, devices and access technologies on user behavior and experience  is discussed in \cite{finamore2011youtube}. Distribution of YouTube's cache servers and their selection process was studied in \cite{adhikari2011you}. 

Our work aims at active measurements and is thus relevant to the LMAP and IPPM WGs. LMAP provides a framework for large scale measurements\cite{draft-lmap-fw}. The model we propose for large scale video performance measurements in Section \ref{lmapmodel} builds upon the LMAP framework. The testing is to be carried out from an LMAP measurement agent (MA).  Regular (long-term) active measurements add additional traffic on the network and should run during idle or low user activity periods, so that they do not interfere with other traffic. Therefore, both the traffic generated and the extra traffic lasts should be minimized.  This implies that we cannot run active measurements for tens of minutes or hours to measure performance of a long video. \cite{morton2013advanced} emphasizes the need for stronger descriptions for test streams because of the indeterministic nature of the Internet. Since we are proposing active measurements using online services, we will be using a variety of test streams, and it is important that they are characterized. In this paper, we propose possible methods to achieve this.

\section{Datasets} 
\label{dataset}

We present results from three measurement activities that we did for YouTube during 2013-2014. The datasets constitute the list of video
URLS extracted from the YouTube's chart pages\footnote{www.youtube.com/charts.YouTube has
changed their service since the collection of the data and now redirects requests for these
pages to a YouTube channel.} for 58 different locations and with popularity defined for differing time
periods (today, last week, last month and all time). 

Our first measurements were based on the charts of July 5, 2013 and we collected the description of the videos available on the YouTube page, including the title, category, number of views, likes and dislikes, available formats, resolutions and file sizes. We collected the charts again for September 11, 2013 and, for this set, in addition to the descriptions as above, we also gathered the date of uploads and file sizes for some selected formats and resolutions. After removing redundancies, there
are over 75,000 videos in each of these datasets and over 130,000 unique videos altogether. About 28\% of the videos of the
July dataset are also present in the dataset of September, of which 85\% are from the all time
charts. At the time, YouTube did not provide any support for Dynamic
Adaptive HTTP Streaming (DASH) and the data collected was for non-adaptive videos.

YouTube introduced the DASH format\footnote{DASH support is seen in
\emph{``stats for nerds''} on YouTube} in October 2013. The DASH
implementation uses a fragmented MP4 file in which the stream is divided into
subsegments for easy switching between different resolutions. Currently, the
subsegment duration used by YouTube as per our dataset is 5 seconds for video
tracks. While YouTube provides audio and video tracks in a single file for
progressive download, the DASH streams provides them in separate files.
To characterize the variation in instantaneous bit rate (burstiness) of video,
we collected information about the frame sizes and timestamps for each video
into a separate dataset (\emph{Frame-logs}), which was collected in April 2014 and so includes both DASH and non-adaptive video streams. We collected logs for MP4 videos in 360p and 720p resolutions, and DASH MP4 videos with 360p, 480p, 720p and 1080p resolutions. Table~\ref{tab_ds} shows a
summary of the sizes of these datasets.

The data were collected at Aalto University using our
YouTube client, designed to run active measurements for YouTube videos. The client is designed to measure YouTube performance for end-users using the \emph{SamKnows} whitebox. It uses \texttt{libcurl}\footnote{http://libcurl.org/} for
fetching videos and extracts YouTube's video metadata using regular expressions by finding keywords in the first HTTP response. The scope of this paper is to characterize videos to aid in designing better measurement systems and not to actually measure network performance. Therefore, we present no results about the quality of the download. YouTube uses numeric identifiers called itags for identifying the formats and resolutions of the video. The itags used during this study are listed in Table \ref{tab_itag}. When collecting \emph{Frame-logs}, the client is run without a rate adaptation algorithm to collect complete frame information for a single bit rate stream.

\begin{table}[!t]
\centering{
\begin{tabular}{lcccc} \hline
	&\textbf{Jul`13}&\textbf{Sep`13}& \multicolumn{2}{c}{\textbf{Frame-logs (Apr `14)}} \\
	&  &  & Non-adaptive & DASH \\ \hline
Videos & 75847 & 76065 & 61269 &33523 \\ \hline
\end{tabular}
}
\caption{Number of videos in the datasets.}
\label{tab_ds}
\end{table}

\begin{table}[!t]
\centering{
\scalebox{0.8}{
\begin{tabular}{ccccc} \hline
\textbf{ITAG} & \textbf{Resolution} & \textbf{Format} & \textbf{Video} & \textbf{Audio}\\ 
& & & \textbf{Codec} & \textbf{Codec} \\ \hline
46&1080p & WebM	& VP8 & Vorbis\\ 
45&720p	 & WebM	& VP8 & Vorbis\\ 
43&360p	 & WebM	& VP8 & Vorbis\\ \hline
37&1080p & MP4  & H264 & AAC\\ 
22&720p & MP4  & H264 & AAC\\ 
18&360p	 & MP4  & H264 & AAC\\ \hline
137&1080p & MP4 DASH & H264 & -\\ 
136&720p & MP4 DASH & H264& -\\ 
135&480p & MP4 DASH & H264& -\\ 
134&360p & MP4  & H264 & -\\ \hline
34&360p	 & FLV  & H264 & AAC\\ \hline
\end{tabular}
}}
\caption{ITAG values of YouTube videos discussed in this paper.}
\label{tab_itag}
\end{table}

\section{YouTube Characteristics}
\label{analysis1}

In this section, we present the analysis of the datasets.

\subsection{Categories}

\begin{figure}[!t]
\centerline {
\includegraphics[width=1.\columnwidth]{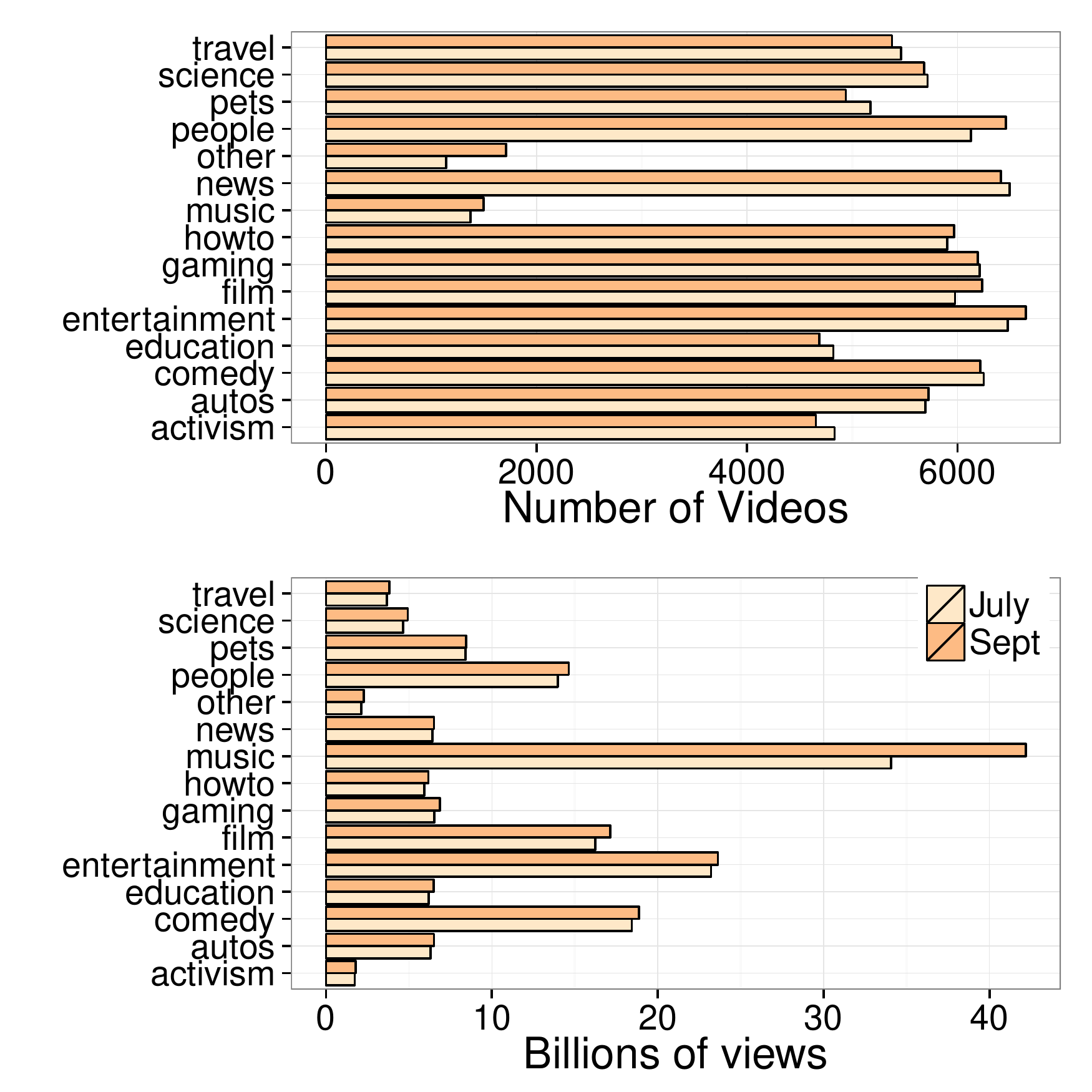}
}
\caption{Distribution of the videos in the July and September dataset and 
	the number of views for each categories. }
\label{fig_cats}
\end{figure}

The datasets cover all the categories of YouTube fairly well, it gives
a good idea of the different types of popular videos available on YouTube.
Figure~\ref{fig_cats} shows the distribution of the video categories and also
the cumulative views for each category. The \textit{Music} category has the
highest number of views despite that less than 2\% of the videos in the
datasets belong to this category. Illegally shared videos are quickly removed,
hence most of the music videos on YouTube are shared by music companies through
syndication hubs~\cite{edmond2012here}. Currently, the most viewed video on
YouTube also belongs to the `music' category. The lower number of music videos also indicates that, unlike other categories, many of the same music videos are popular across multiple countries, resulting in common results for various locations. 

\begin{figure}[!t]
\centerline {
\includegraphics[width=0.9\columnwidth]{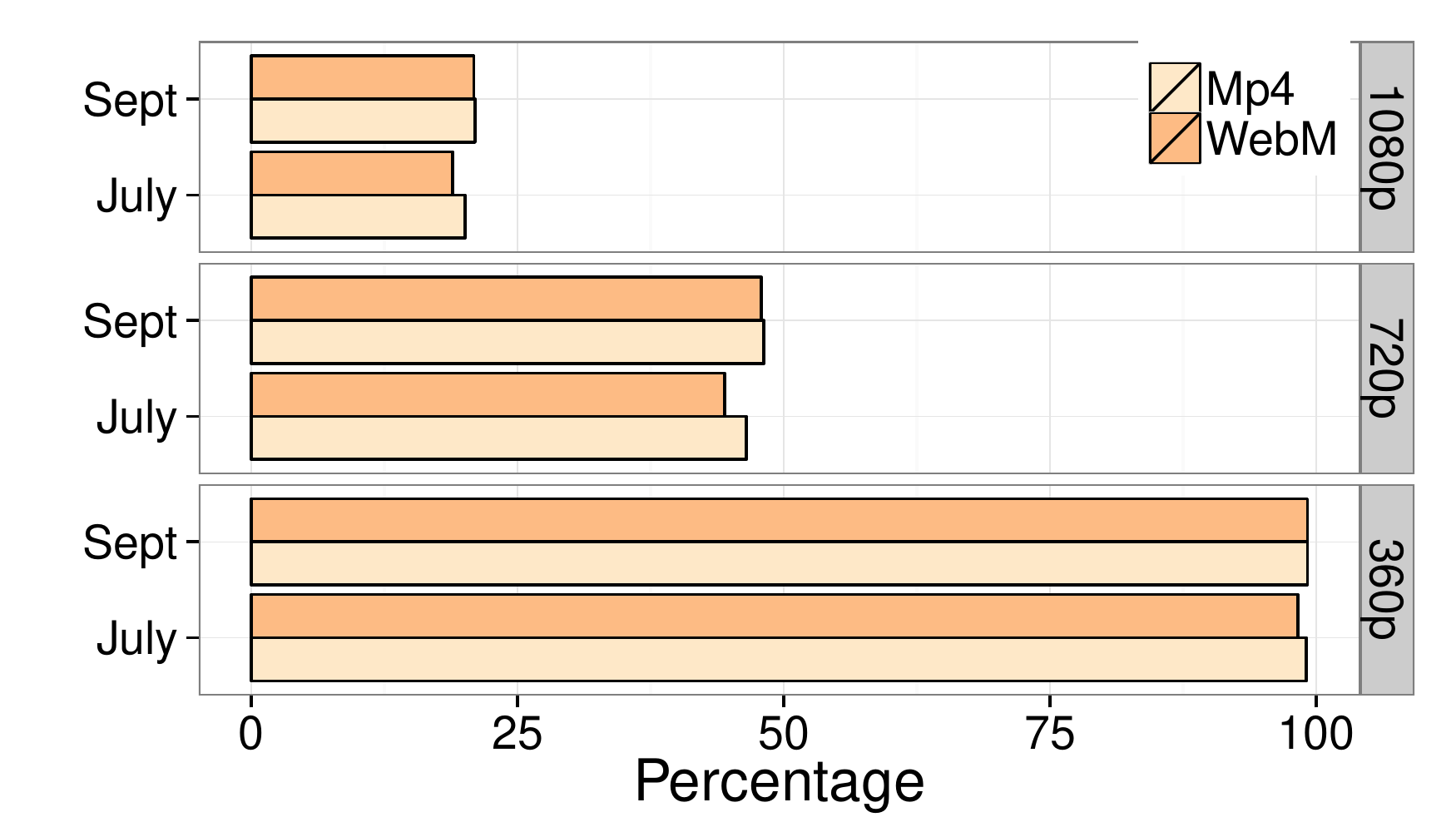}
}
\caption{Availability of WebM and MP4 formats for different resolutions. The July dataset is shown as \emph{ds1} and the September dataset is shown as \emph{ds2}}
\label{fig_formats}
\end{figure}

\subsection{Formats and Resolutions}

Internet video is viewed on a number of different devices and hence a range of different resolutions are supported for compatibility reasons. Furthermore, new resolutions appear and old ones are discarded. Currently, YouTube uses a 16:9 aspect ratio for wide screen displays, and provides videos in 7 different resolutions. In our datasets we observed resolutions as low as 144x176 (QCIF) to as high as 3072x4096 (4K), however YouTube keeps changing the offered resolutions based on technological needs or internal reasons.

We observed that the default resolution of 360p is available
for over 99\% videos in the datasets. If a video is available in MP4 for a
particular resolution, it is also available in WebM for the same resolution.
Less than 1\% videos are available in only one of the two
formats. The overall availability of MP4 format is slightly higher in
comparison to WebM in July, but this gap is not seen anymore in the
September dataset (See Figure~\ref{fig_formats}). 

When YouTube introduced DASH in Oct 2013, it stopped providing non-adaptive streams for Full HD videos. Consequently, the DASH format uses only the fragmented MP4 format and support for WebM was no longer available. The 360p and 480p versions of FLV have been discontinued as well, however 240p is still available.

\subsection{Date of Upload}

\begin{figure}[!t]
\centerline {
\includegraphics[width=0.9\columnwidth]{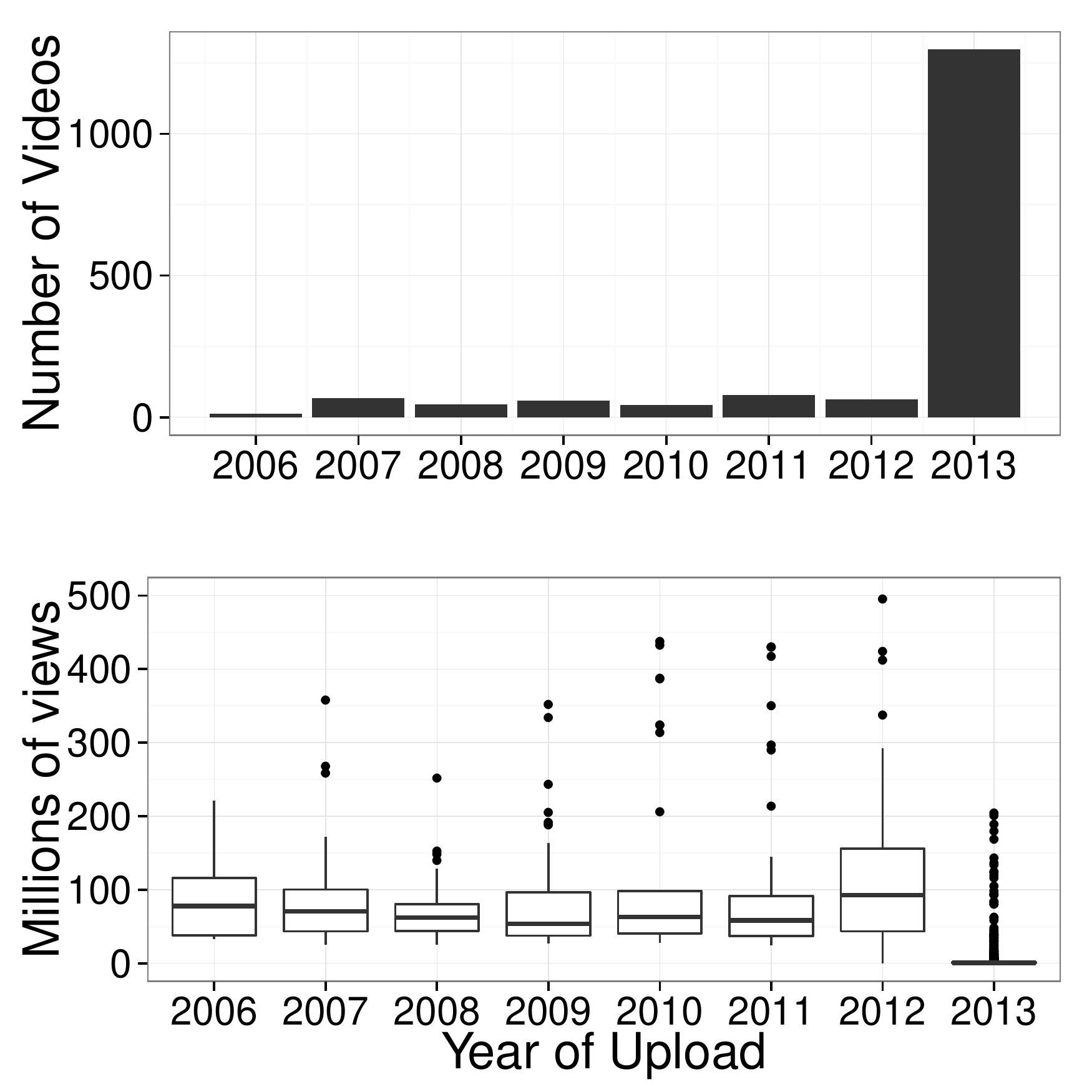}
}
\caption{A chart of the number of videos and box plots of the number of views for these videos based on the year of upload. This data is only for the worldwide
charts of September 11, 2013. The whiskers of the box plots extend to 1.5*IQR (Inter Quartile Range). }
\label{fig_year}
\end{figure}

The date of upload is available only for videos collected in September, where over 72\%
of the videos were uploaded in 2013. The popularity of videos in reference to how long
they have been available is shown in Figure~\ref{fig_year}.  It shows per-year
distribution of videos and the boxplots for the number of views.  The graphs are
based only on the worldwide charts and for the sake of clarity, outliers with more
than 500\,M views are not shown.

\subsection{Video Length}
\label{sec.char.vidlen}

\begin{figure}[!t]
\centerline {
\includegraphics[width=0.8\columnwidth]{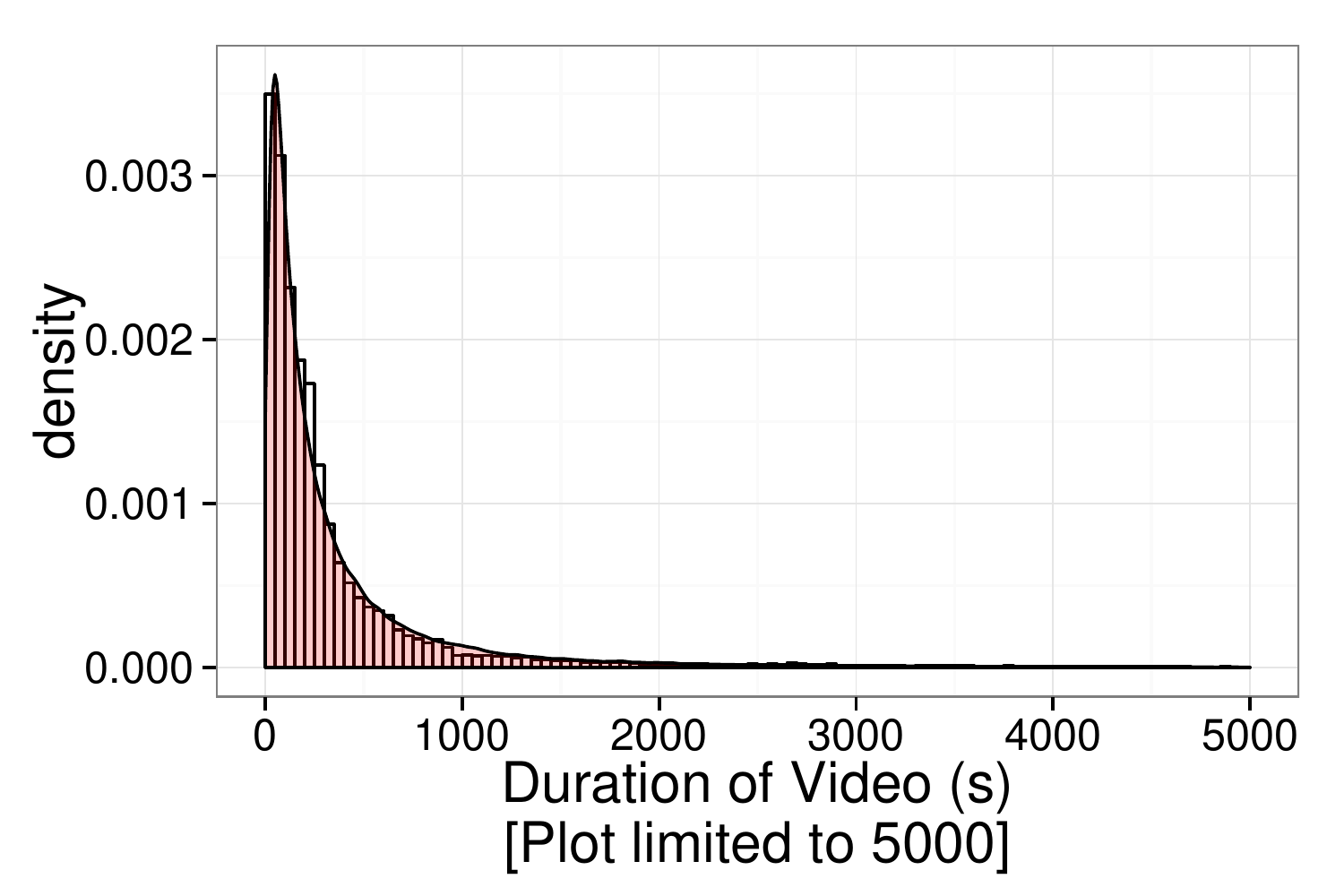}
}
\caption{Distribution Fitting - Histogram of length of the video in seconds and
density plot of lognormal distribution.}
\label{figure_histvdolen}
\end{figure}

\begin{figure}[!t]
\centerline {
\includegraphics[width=0.9\columnwidth]{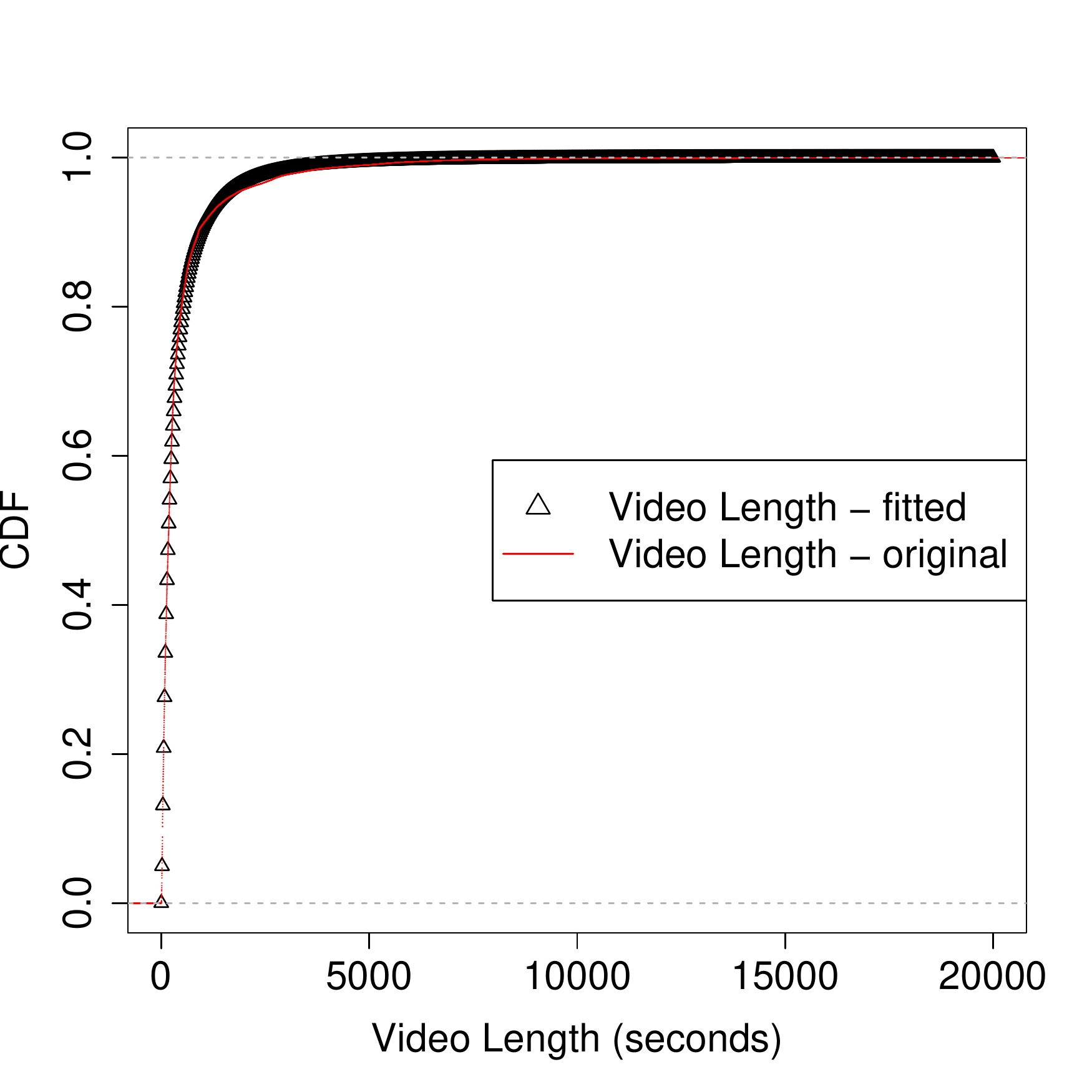}
}
\caption{Distribution Fitting - CDF graphs for Video length and lognormal
distribution. The actual data has a large number of points which makes it appear as a continuous line, the fitted distribution is shown as points to make it more visible over the actual. }
\label{figure_ecdfvdolen}
\end{figure}

The longest YouTube video in the datasets is over 11 hours long. Since YouTube
allows some users to upload 10 hour long videos, there are a number of videos
that last for more than an hour. The average duration of the videos in the
complete dataset is \emph{441s} and the median is \emph{181s}. The video length
fits a lognormal distribution but the tail is heavily skewed with only 15\%
of the videos with durations longer than \emph{600s}. 50\% of the videos in the Inter-quartile Range (IQR)
have durations between \emph{71} and \emph{387} seconds. We suggest that this is a good range for active measurements, as the videos are long enough to gather interesting results and yet not so long that the extra traffic starts interfering with user traffic.   Figures~\ref{figure_histvdolen} and \ref{figure_ecdfvdolen} illustrate the suitability of the lognormal fit with an empirical cumulative distribution function (CDF) plot and, an histogram-density plot, respectively.

\begin{figure}[!t]
\centerline {
\includegraphics[width=0.75\columnwidth]{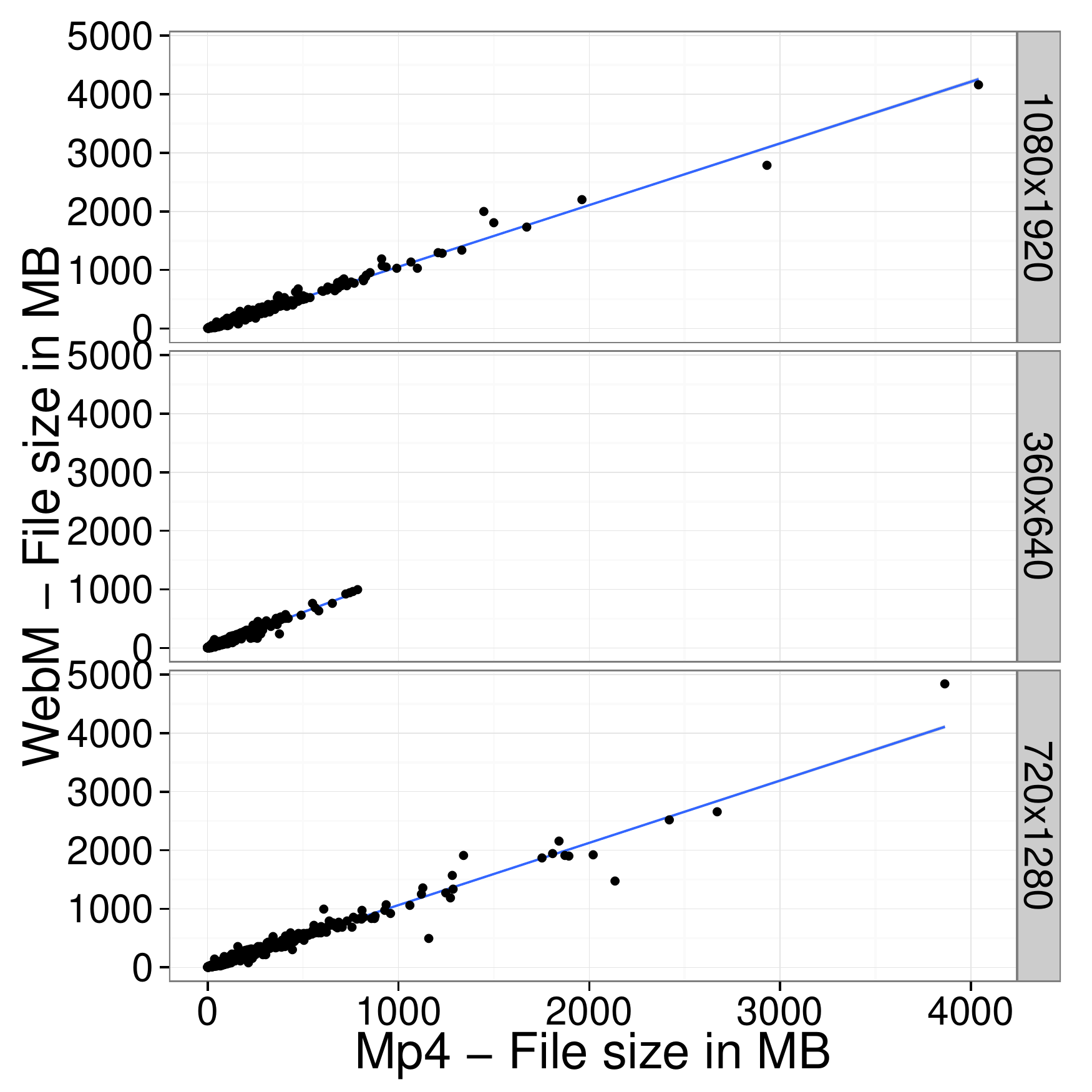}
}
\caption{Correlation in the size of an MP4 file and a WebM file for a given
resolution. The slopes of the lines are slightly above 1, indicating that WebM files are generally larger than MP4 files.}
\label{fig_corrsizes}
\end{figure}

\begin{figure}[!t]
\centerline {
\includegraphics[width=0.9\columnwidth]{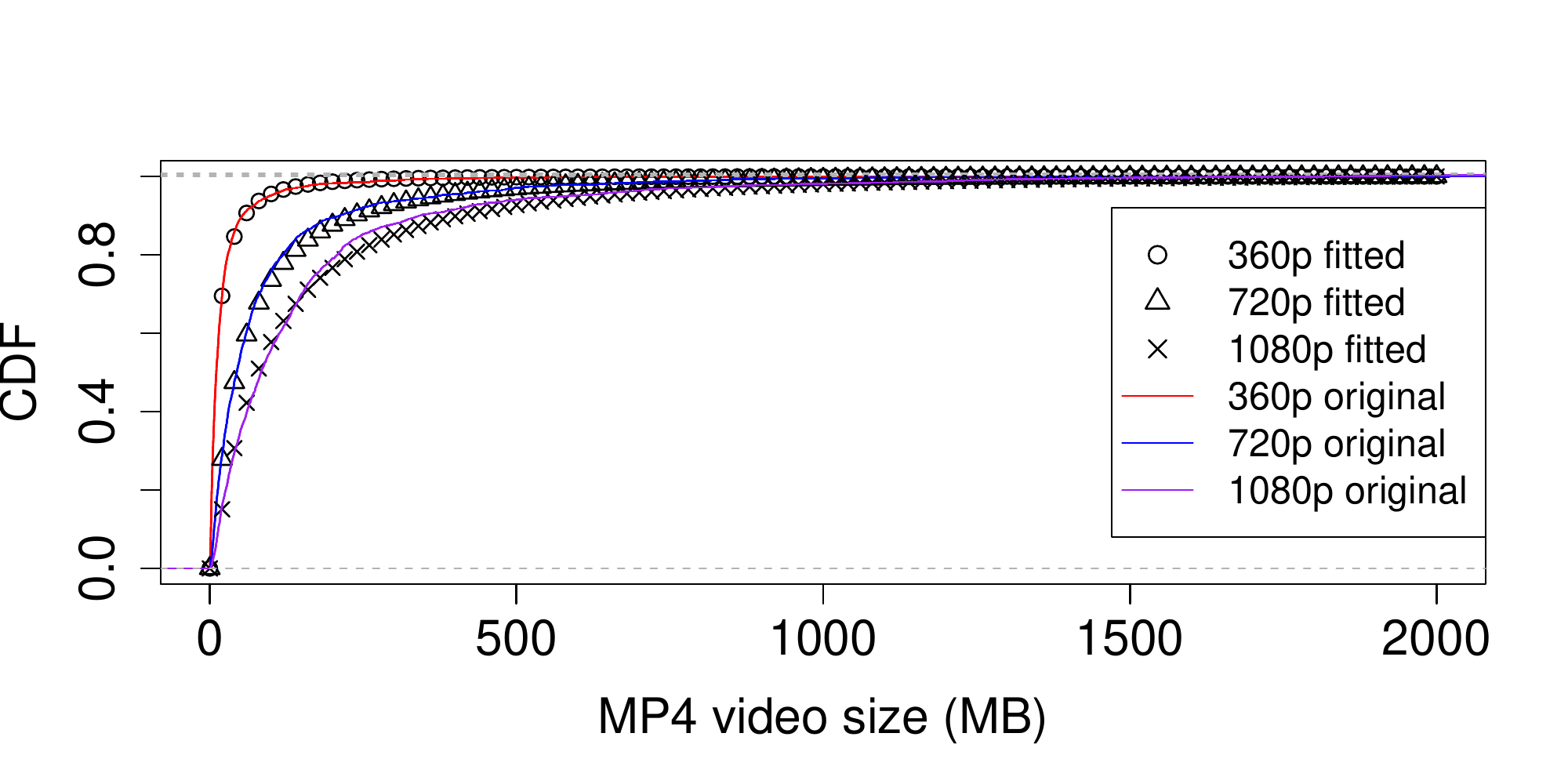}
}
\caption{Distribution Fitting - MP4 Video Sizes (MB) CDF. The actual data has a large number of points which makes it appear as a continuous line, the fitted distribution is shown as points to make it more visible over the actual. }
\label{fig_ecdfvdos}
\end{figure}

\begin{figure}[!t]
\centerline {
\includegraphics[width=\columnwidth]{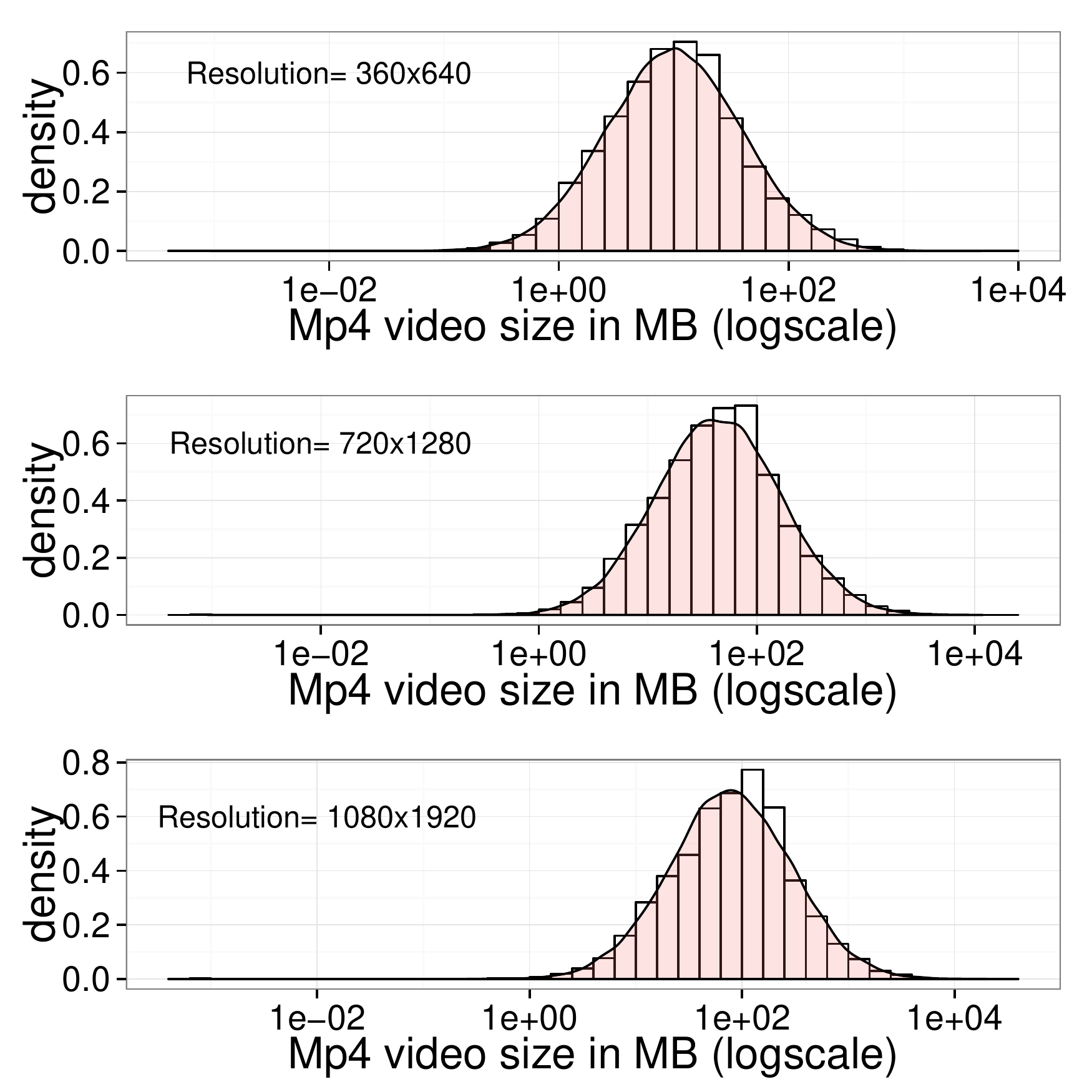}
}
\caption{Distribution Fitting - MP4 Video Sizes (MB) histograms with lognormal
density curves}
\label{fig_histvdos}
\end{figure}

\begin{table*}[!t]
\centering{
\scalebox{0.9}{
\begin{tabular}{p{2cm}cccccp{2.5cm}} \hline
& \textbf{Sample size}& \textbf{Median} & \textbf{Mean} & \textbf{Max} & \textbf{Std Dev} & \textbf{Distribution Fit} \newline (lognormal) \newline (meanlog, sdlog)\\ \hline
MP4 360p&75582&10.18&25.74&2074&58.20&(2.30, 1.35)\\ 		
MP4 720p&36703&45.48&103.8&6991&208.73&(3.77, 1.32)\\ 
MP4 1080p&15913&85.66&150.61&6700&325.70&(4.36, 1.32)\\ 
WebM 360p&75576&11.53&23.378&2619&71.04&(2.42, 1.42)\\ 
WebM 720p&36322&47.63&89.06&8122&225.09&(3.80, 1.37)\\ 
WebM 1080p&15819&89.68&160.96&11050&364.42&(4.39, 1.35)\\  \hline
\end{tabular}
}}
\caption{Distribution fitting for File sizes (MB) for different formats. Analysis based on data collected before YouTube introduced DASH MP4. The 1080p analysis is based on the non-adaptive files for each format, which are no longer available; only DASH videos are available for 1080p resolution on YouTube now.  }
\label{table_fs}
\end{table*}

\begin{table*}[!t]
\centering{
\scalebox{0.9}{
\begin{tabular}{p{2.5cm}cccccp{2.5cm}} \hline
& \textbf{Sample size}& \textbf{Median} & \textbf{Mean} & \textbf{Max} & \textbf{Std Dev} & \textbf{Distribution Fit} \newline (lognormal) \newline (meanlog, sdlog)\\ \hline
Video length (s)&132272&181&315&42400&1025&(5.16, 1.31)\\ \hline
\end{tabular}
}}
\caption{Distribution fitting for Video Length (s) using unique videos from both datasets of July and September}
\label{table_len}
\end{table*}

\subsection{File Sizes}
\label{sec_filesizes}

For videos that are available in both MP4 and WebM formats, the sizes of the
files in each format for a particular resolution are comparable with some
exceptions. Figure~\ref{fig_corrsizes} shows the correlation for all three
resolutions, we observe 0.99 correlation. The best fit for the file sizes is a
lognormal distribution. Table~\ref{table_fs} shows a summary of distribution
fits for each resolution and format for September measurements. The table also shows the
sample size used for the fitting depending on the number of videos available
for that format and resolution. The actual data has a long skewed tail due to
some videos with duration well over an hour, which causes it to deviate from
lognormal. Figures~\ref{fig_ecdfvdos} and \ref{fig_histvdos} illustrate the
suitability of the lognormal fit with a histogram-density graph and CDF.

\begin{figure}[!t]
\centerline {
\includegraphics[width=0.9\columnwidth]{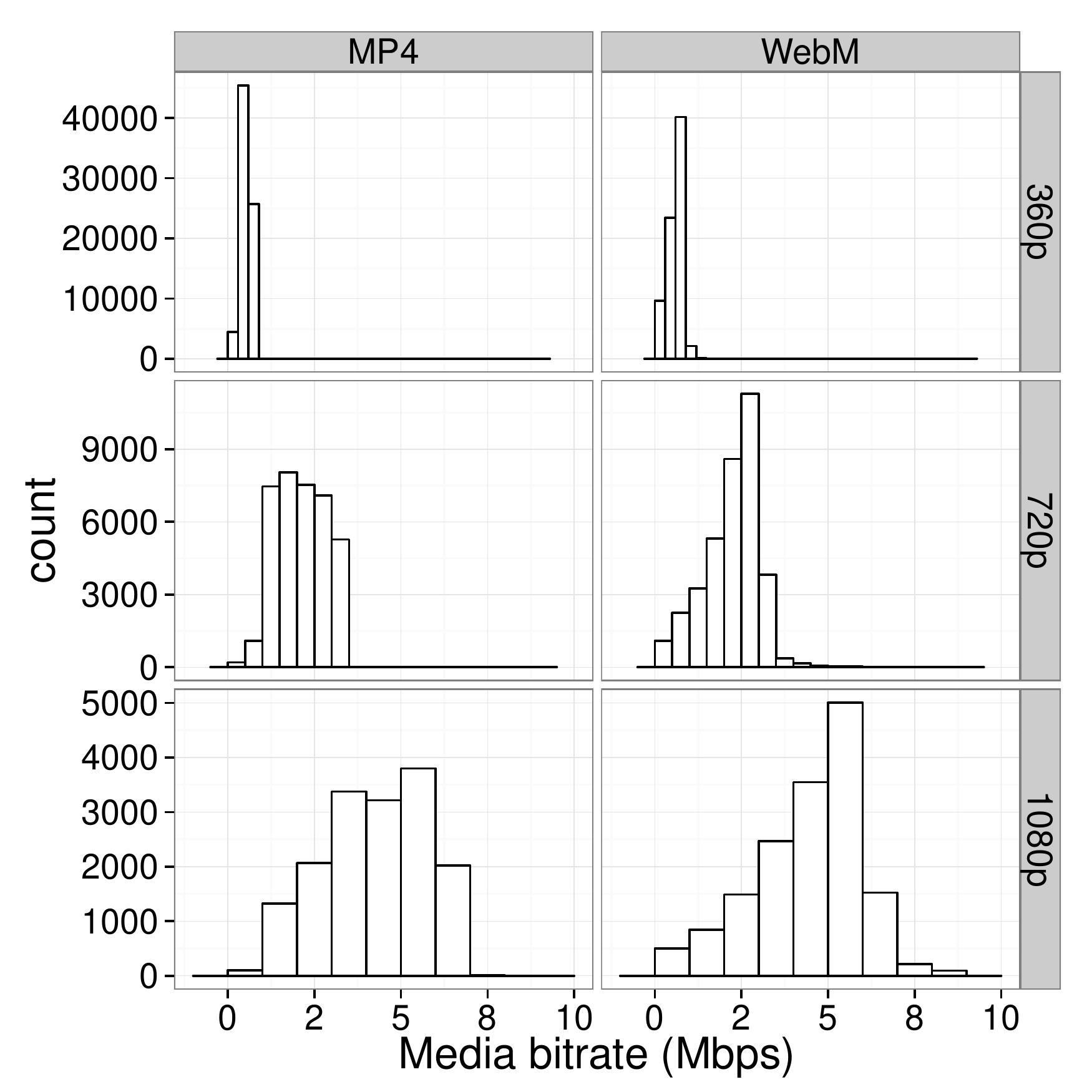}
}
\caption{Average bit rate of non-adaptive MP4 and WebM calculated by dividing file sizes by the video durations. Note that these are not the encoding bit rates for the videos, since they include both metadata and the audio streams. WebM videos are generally larger than MP4 as was shown in Figure \ref{fig_corrsizes} and hence the bit rates for WebM are also higher than those for MP4.}
\label{fig_ds12bitrates}
\end{figure}

\subsection{Media Bit Rates}
\label{sec.char.mr}

Media is encoded using variable encoding bit rates, which means the bit rates
can spike to values much higher than the advertised bit rate for a media
stream. Furthermore, depending on the content, different video or audio
streams encoded using similar encoding parameters may have different resulting
average bit rates. We calculated the average bit rates for videos from the
September  dataset by dividing the file sizes by the duration of the
corresponding video. The results are shown in Figure~\ref{fig_ds12bitrates}.
WebM files are generally larger as we showed in Section \ref{sec_filesizes},
and hence WebM bit rates are higher than MP4 for similar resolutions as well.

\begin{figure}[!t]
\centerline {
\includegraphics[width=0.75\columnwidth]{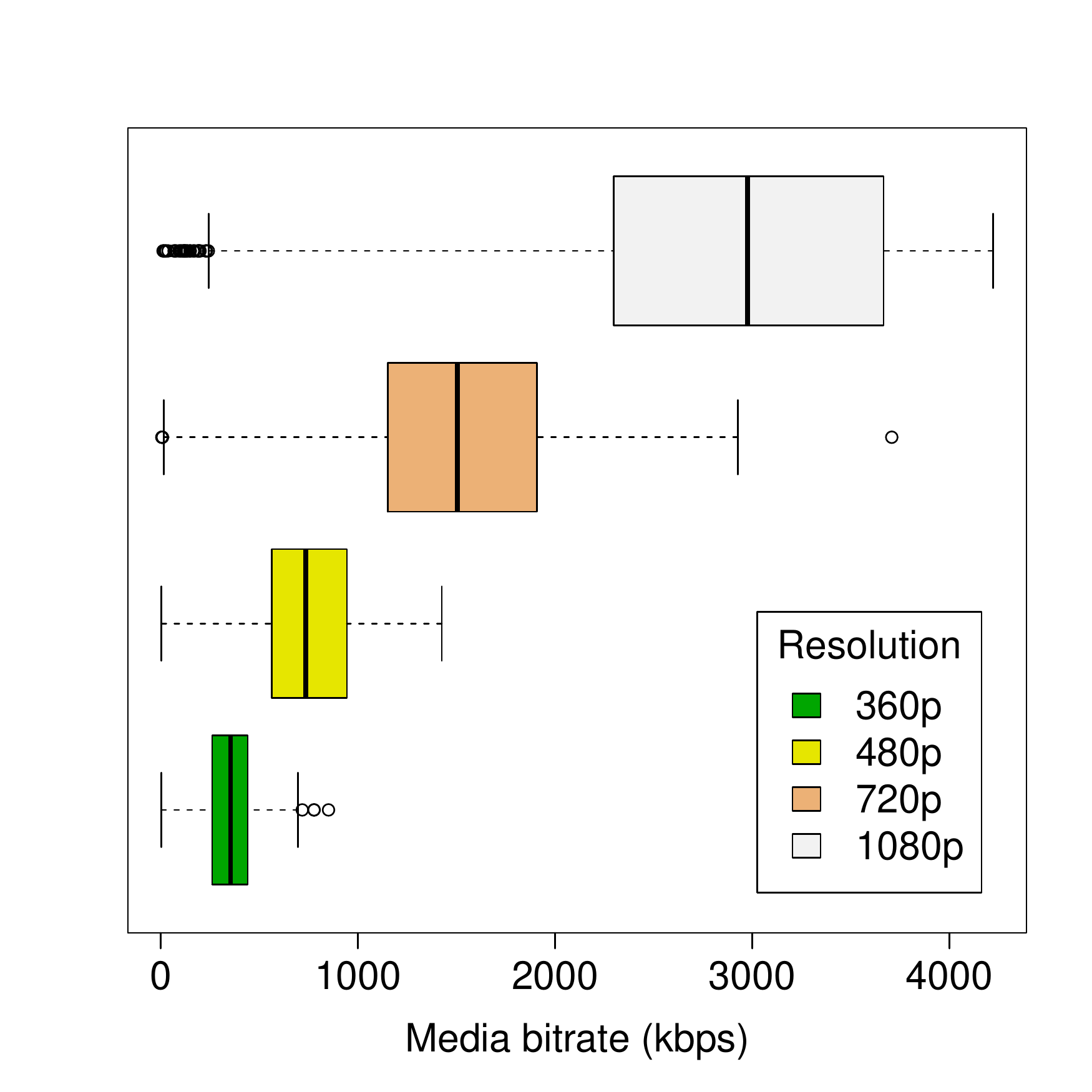}
}
\caption{Average bit rate of DASH MP4 calculated for the first 3 minutes or the entire duration for files under 3 minutes}
\label{fig_fragmean}
\end{figure}

\begin{figure}[!t]
\centerline {
\includegraphics[width=0.6\columnwidth]{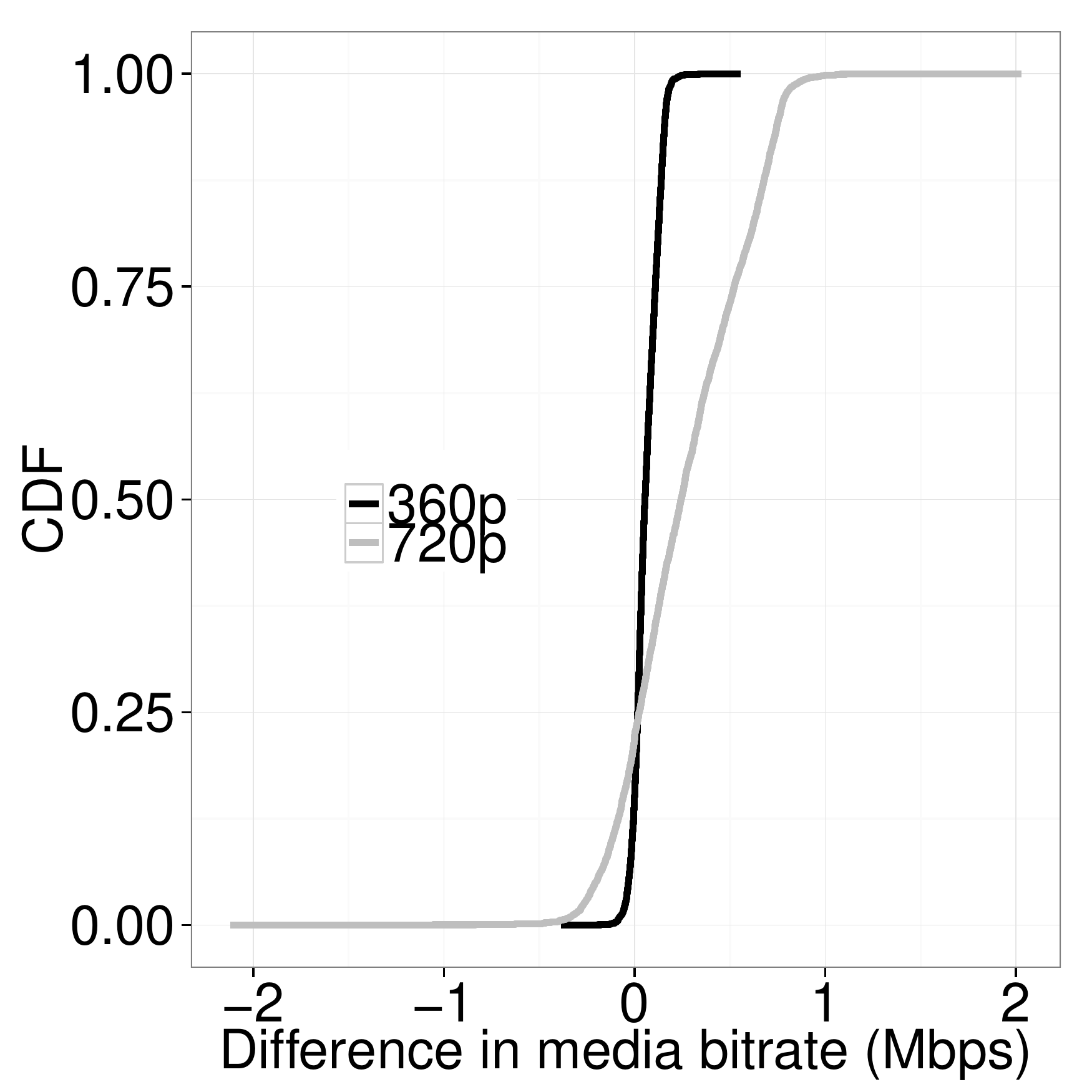}
}
\caption{The average bit rate of non-adaptive MP4 video minus the average bit rate of DASH MP4 video, calculated for first 3 minutes. For most files, the DASH videos have a lower bit rate. These bit rates are calculated using video frames and hence represent video encoding bit rates. Audio frames and metadata are not included. }
\label{fig_fragvsunfragbitrate}
\end{figure}

For the DASH videos in our dataset, we had frame level information for up to three minutes of the video (audio files are separate for DASH, these calculations are based on video frames only). We calculated average bit rates for different resolutions by summing the frame sizes and dividing by the total duration. Figure \ref{fig_fragmean} shows the summary of the bit rates. We found the adaptive streams of 360p and 720p to have lower average bit rates in comparison to the non-adaptive streams as shown in Figure \ref{fig_fragvsunfragbitrate}. YouTube stopped serving non-adaptive file formats for 480p and 1080p when it introduced DASH. Our DASH analysis takes only 3 minutes of video into account; we will show later in Section \ref{sec_3min} that this results in little loss of information, if any. 

\subsection{Burstiness}
\begin{figure}[!t]
\centerline {
\includegraphics[width=0.9\columnwidth]{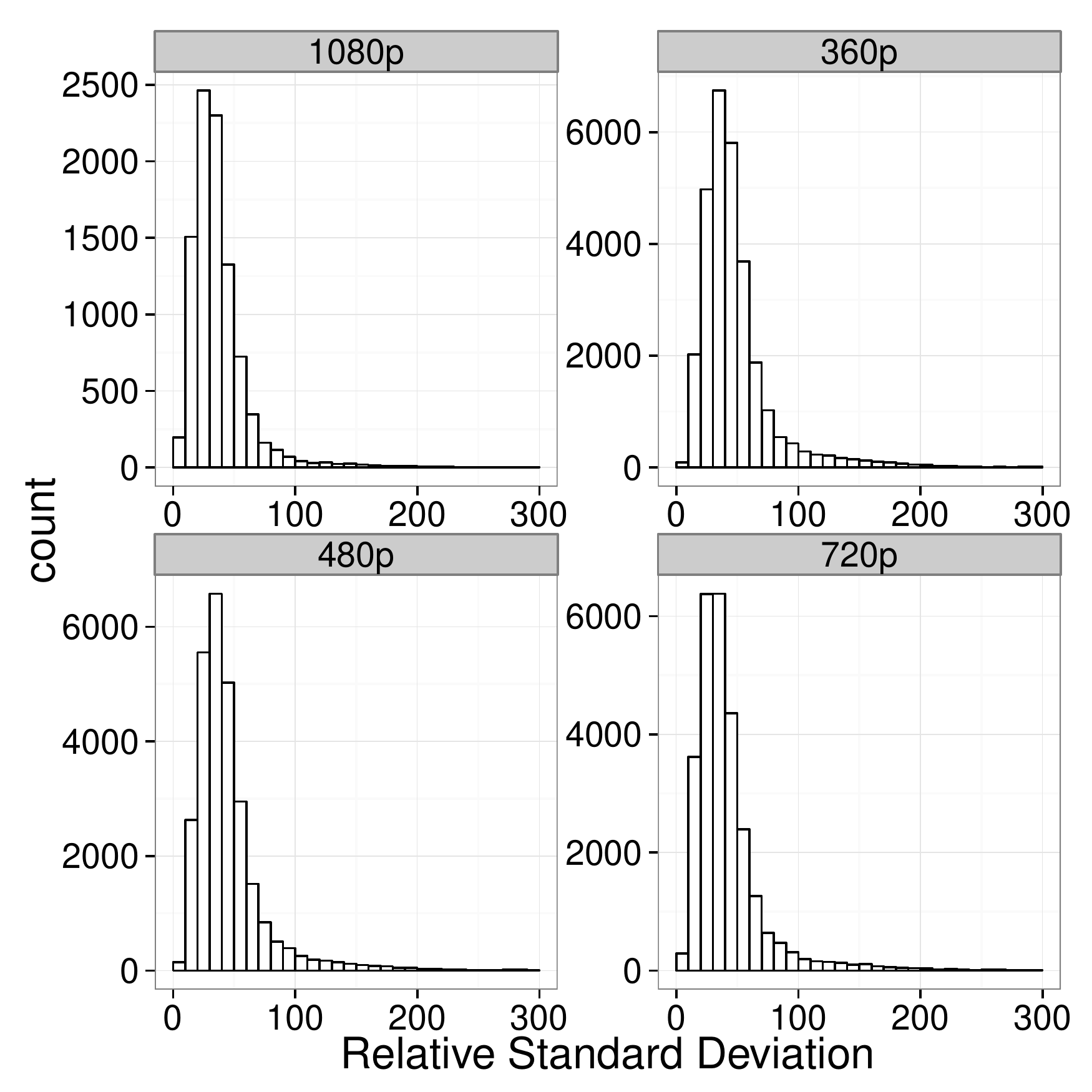}
}
\caption{Relative standard deviation in 1 second instantaneous bit rate of DASH videos for different resolutions. Note: Scales are different. All resolutions are for the same set of videos, however, due to unavailability of higher resolutions for some videos, the counts are different.  }
\label{fig_fragsd}
\end{figure}

A video with sudden traffic bursts is more likely to cause a freeze in the playout than one that has a more consistent rate. We measured the instantaneous bit rates as the bit rate observed during one second of playout. The burstiness of the video can be related to the standard deviation in these instantaneous bit rates. We can use the mean bit rate and this value of burstiness to classify Internet video into groups. This can aid in comparing results of active measurements done over a range of different videos. Both these parameters are easy to measure at a measurement agent during the test and can be recorded along with other results. Gathering this information will also help in keeping up-to-date information about the characteristics of Internet video as a side-effect of measuring performance. Figure \ref{fig_fragsd} shows the relative standard deviation for different resolutions of DASH video for up to 3 minutes of video. 

\begin{figure}[!t]
\centerline {
\includegraphics[width=\columnwidth]{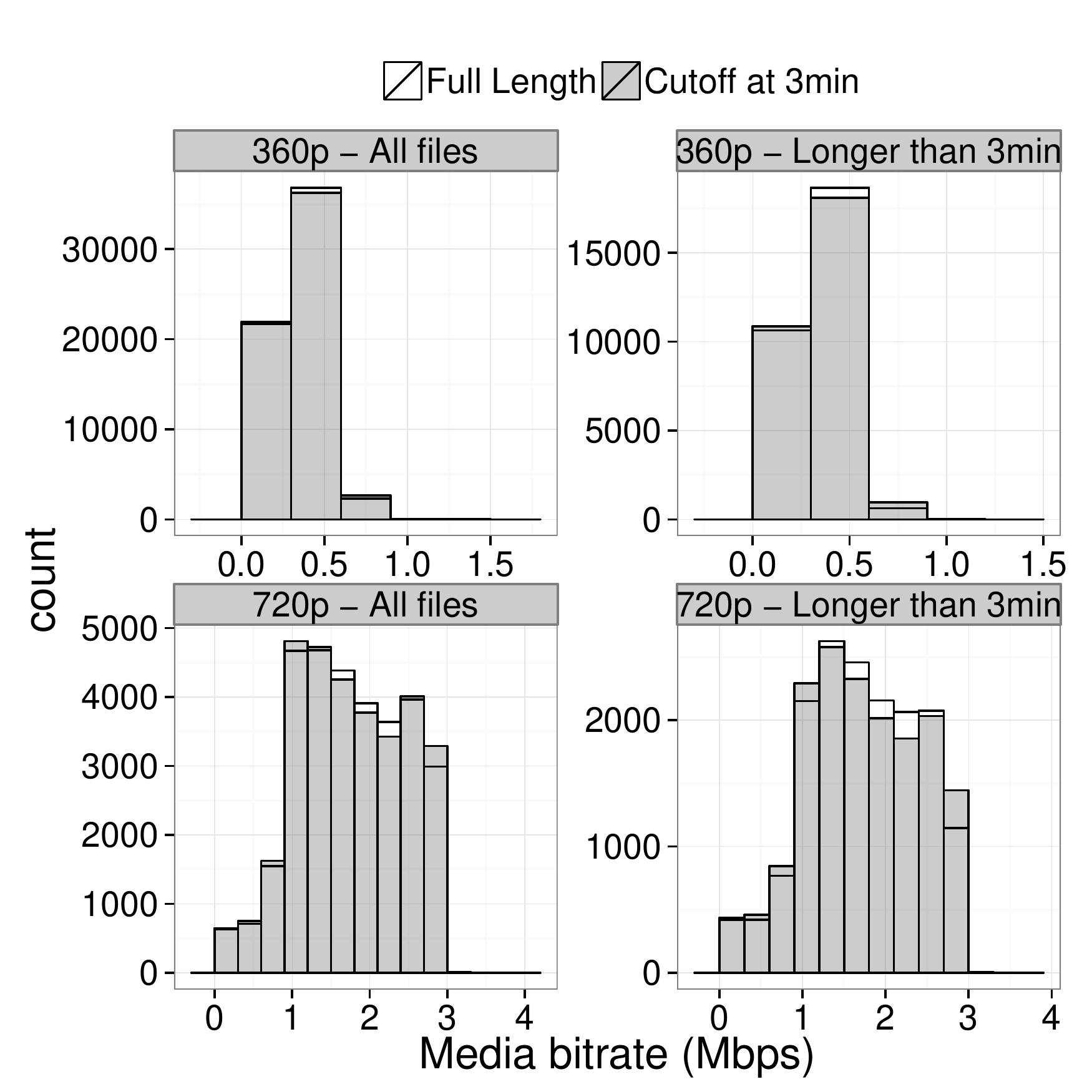}
}
\caption{Distribution of average bit rate of non-adaptive MP4 videos when calculated over the entire duration of the video and when calculated for only the first 3 minutes. Histograms are shown for the whole set of files, and also for only the files with duration greater than 3 minutes as these are the ones that will be affected by the cut-off. The difference is very low in both cases. Note that the scales are not the same for all 4 graphs.  }
\label{fig_unfragmeans}
\end{figure}

\subsection{Three Minutes Cut-off}
\label{sec_3min}
\begin{figure}[!t]
\centering {
\includegraphics[width=0.9\columnwidth]{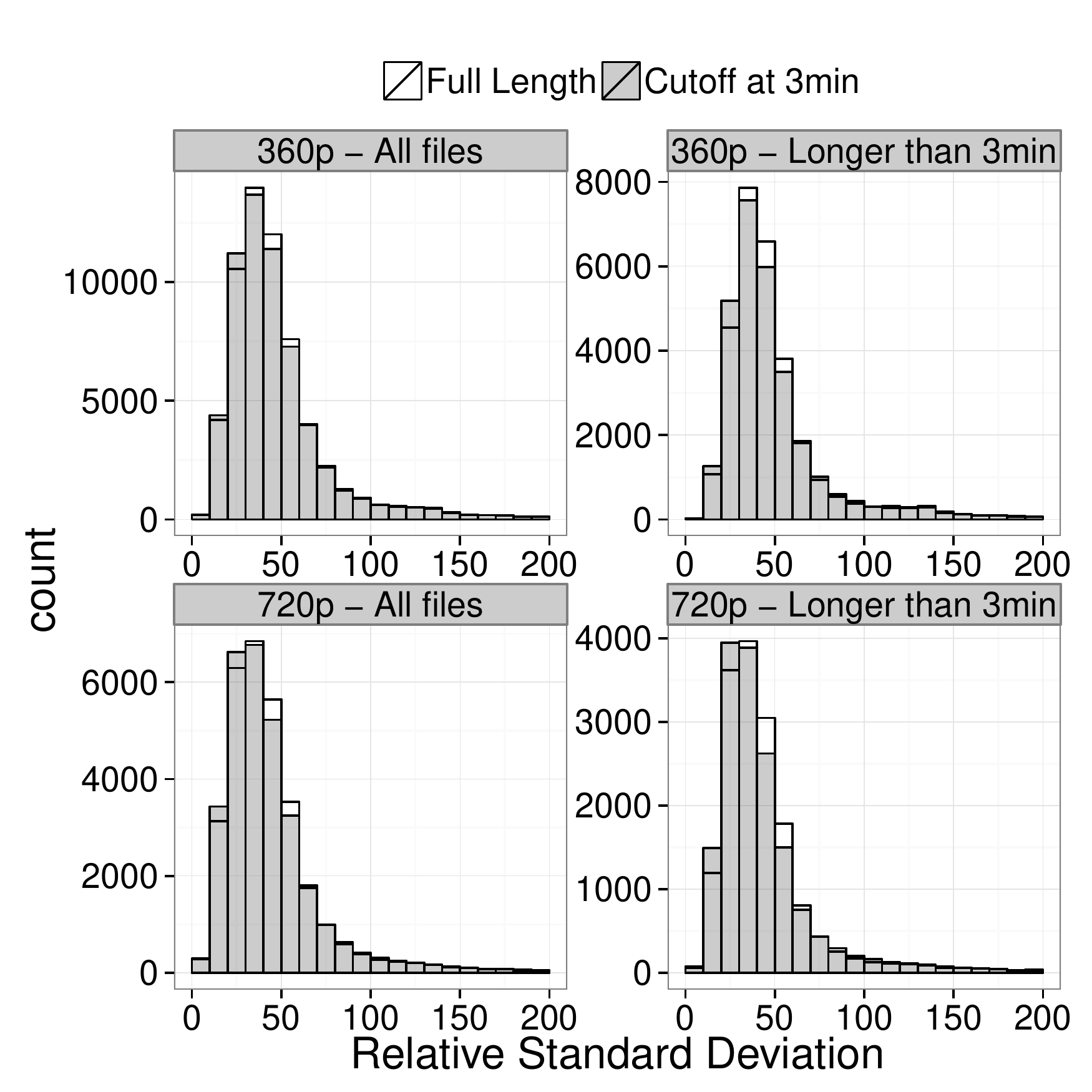}
}
\caption{Average bit rate of non-adaptive MP4 calculated for the first 3 minutes or the entire duration for files under 3 minutes}
\label{fig_unfragsd}
\end{figure}

We calculated the average bit rate of non-adaptive MP4 videos for the first 3 minutes and compared it to the average bit rate of the entire video. As expected, the values are comparable and the distribution of average bit rates remains the same; illustrated in Figure \ref{fig_unfragmeans}. We did the same exercise with the standard deviation in per-second instantaneous bit rates of the videos and found similar results with minor changes in the shape of the histogram as shown in Figure \ref{fig_unfragsd}. Hence, active measurements that span for only 3 minutes can be enough for measuring performance to a good degree of accuracy, at least for short videos like YouTube. It is worth mentioning, that the test does not have to run for the entire playout duration. Once the 3 minutes of video is downloaded, the test can calculate whether or not all frames have arrived before playout time and terminate. 

This result may not apply to long videos such as movies. We discuss this in more detail with suggestions on how to handle exceptions in Section \ref{discussion}.
 

\begin{figure}[!t]
\centerline {
\includegraphics[width=0.95\columnwidth]{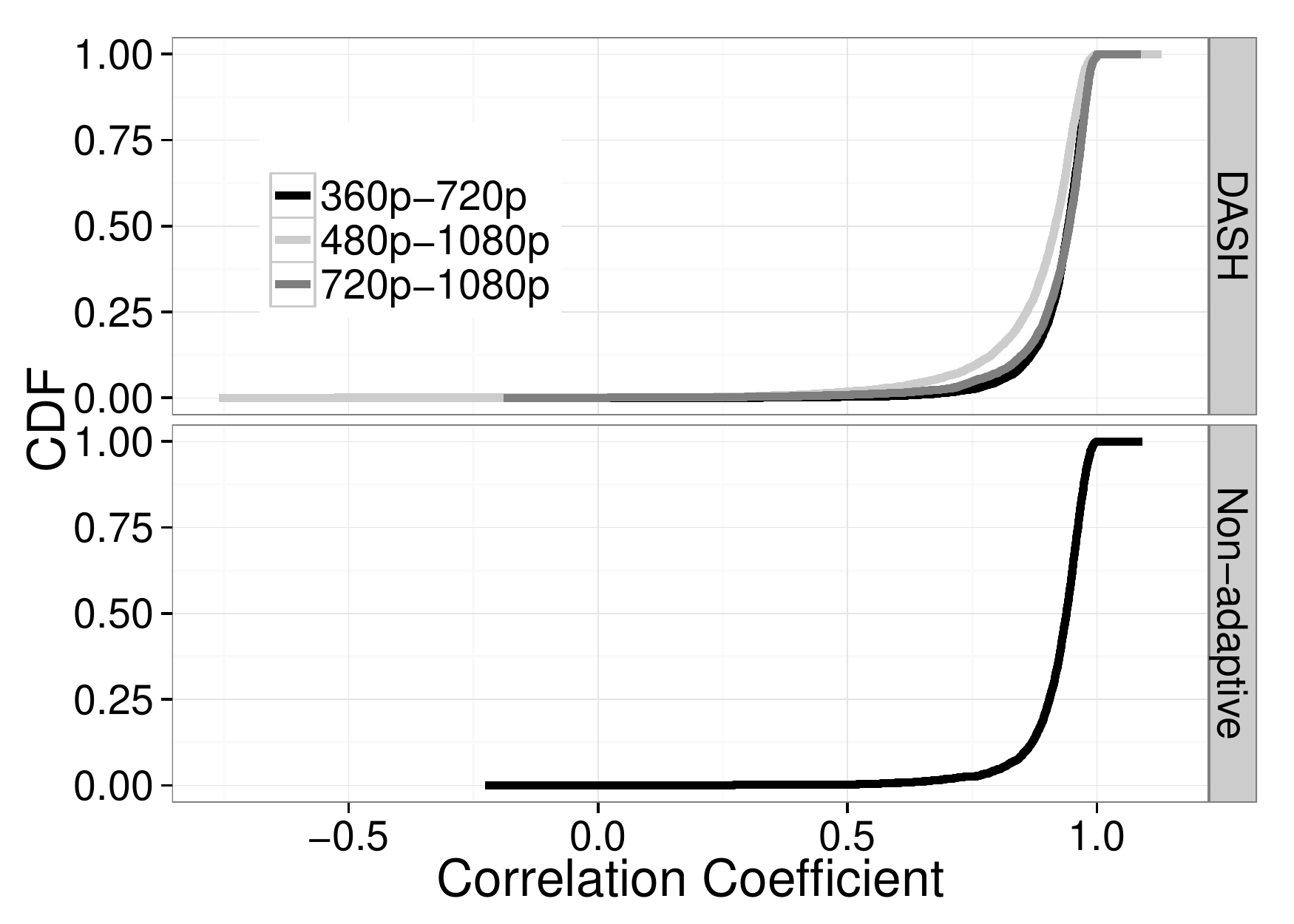}
}
\caption{CDF of correlation coefficients between different resolutions for DASH and the stream of a video at lower resolution with the stream at a higher resolution. The CDFs shown here are made using samples to reduce number of points in the plot, the shape is the same but the tails extend to -1 for the full set of videos, with only a small number of files with negative correlation. }
\label{fig_ecdfcor}
\end{figure}

\section{Simulating Various Resolutions}
\label{analysis2}
We observed that the correlation in the per-second instantaneous bit rates of videos is very strong, as shown in Figure \ref{fig_ecdfcor}, so upscaling or downscaling a video to a higher or lower resolution can be done with a good level of accuracy. This kind of simulation has two use cases: 1) in Active measurement for performance testing from a user's own test server; we can store different files in a single resolution and simulate traffic for other resolutions. 2) For generating traffic for testing DASH algorithms. Such a scheme is useful for saving storage space, since high resolution videos are in the range of Gigabytes and it may not be worth storing so much data when none of it is actually viewed at the other end. It is also useful in cases where you simply might not have access to higher resolution videos to be used for testing. 

We used both DASH and Non-adaptive MP4 videos for testing this hypothesis. We explore three scenarios based on three different use cases: 

\begin{enumerate}
\itemsep0pt \parskip0pt \parsep0pt
  \item Full HD - We simulate 1080p from a 480p video and vice versa using 8045 DASH streams 
  \item Mobile - We simulate 720p from 360p and vice versa using 22,600 DASH streams
  \item Non-Adaptive - We simulate 720p from 360p and vice versa using 30,871 non adaptive streams
\end{enumerate}

\subsection{Methodology}

Our analysis for DASH uses a cut-off value of 3 minutes for the videos. The cut-off would only introduce minimal bias in our results because the correlation coefficients remain almost the same as we saw in Section \ref{sec_3min}. The instantaneous bit rate \(B_{sim_i}\) of a simulated resolution with average bit rate \(\bar{B}_{sim}\) at i-th second, can be calculated from a different resolution with average bit rate \(\bar{B}_{orig}\) and instantaneous bit rate \(B_{orig_i}\) using the formula: 
\[B_{sim_i} = \frac{B_{orig_i} \times \bar{B}_{sim}}{\bar{B}_{orig}}\]

The value of i is a multiple of the measurement interval we take for measuring instantaneous bit rates. We used 1 second intervals.

\subsection {Evaluation}
To compute the error in a simulated video, we used Mean Average Percentage Error (MAPE) calculated using the following formula, where n is the total number of instantaneous bit rate values:
\[MAPE =  \frac{1}{n}\sum\limits_{i=1}^n\frac{ |B_{orig_i}-B_{sim_i}|}{\bar{B}_{orig}} \times 100\]

Table \ref{tab_mape} shows the mean and 95th percentile of the MAPEs for all videos used for mobile, Full HD and non-adaptive case. The slightly higher values when downscaling are a consequence of the unsymmetric nature of MAPE. 

\begin{figure*}[!t]
	\centering{
    \subfloat[ECDF of original 1080p video streams and the simulated 1080p stream created by upscaling 480p stream.\label{subfig-hd:dummy}]{%
      \includegraphics[width=1.8\columnwidth]{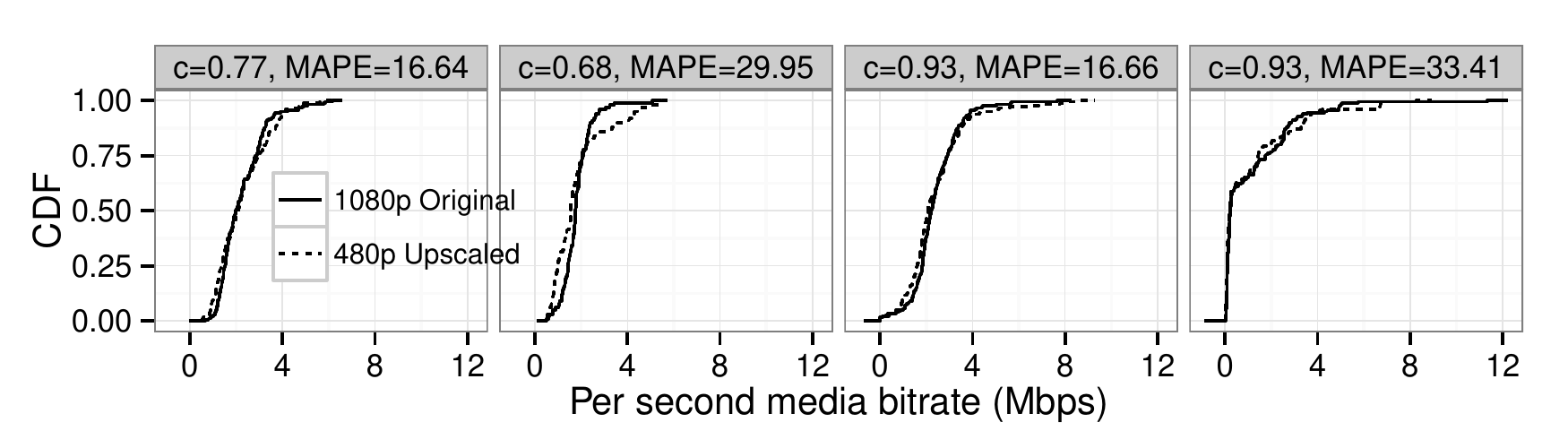}
    }}
    \centering{
    \subfloat[ECDF of original 480p video streams and the simulated 480p stream created by downscaling 1080p stream.\label{subfig-hd:dummy}]{%
      \includegraphics[width=1.8\columnwidth]{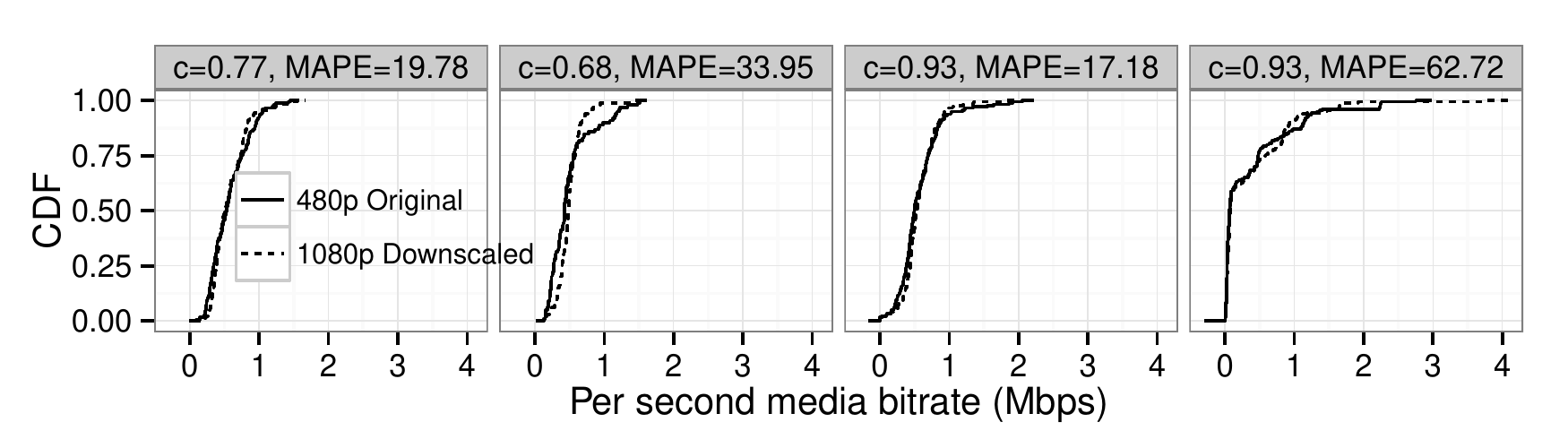}
    }}
    \caption{We picked four videos with different correlation coefficients (c) to show the CDF comparison of a simulated (upscaled or downscaled) stream with the original. The same four videos are represented in (a) and (b). The videos were picked so that they represent a range of c and MAPE values. Notice that the Figure(a) and (b) are in a way inverses of each other. The case with c=0.68 shows a bad fit, as expected; however, such a case is less likely since 90\% of the videos have $c \textgreater 0.75$ for 1080p and 480p}
    \label{fig_ecdfhd}
  \end{figure*}

\begin{figure*}[!t]
	\centering{
    \subfloat[CDF of original 720p video streams and the simulated 720p stream created by upscaling 360p stream.\label{subfig-1:dummy}]{%
      \includegraphics[width=1.8\columnwidth]{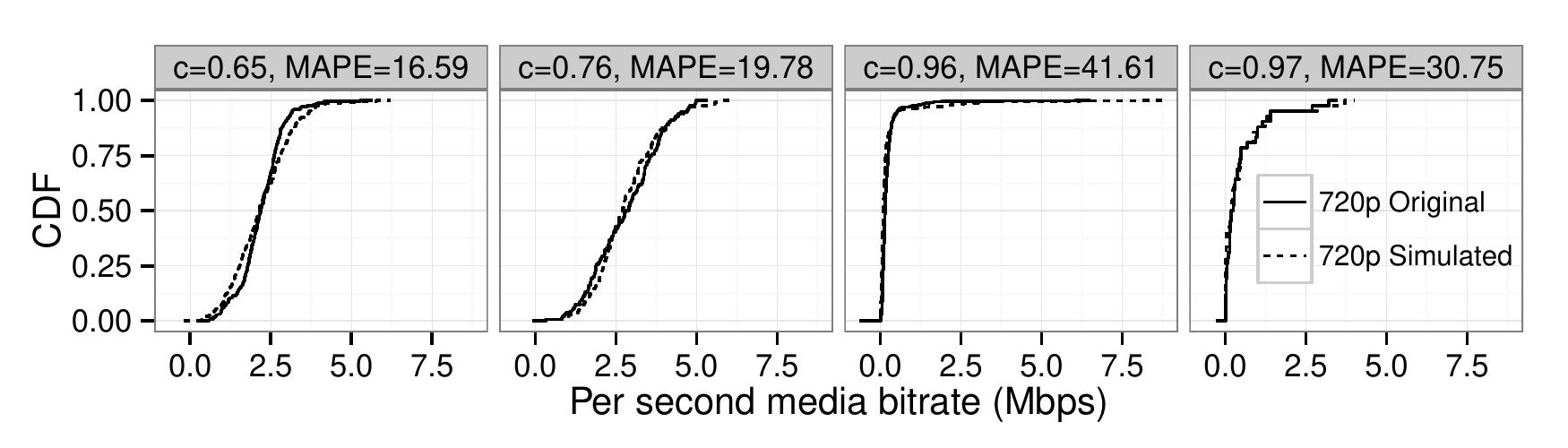}
    }}
    \centering{
    \subfloat[CDF of original 360p video streams and the simulated 360p stream created by downscaling 720p stream.\label{subfig-2:dummy}]{%
      \includegraphics[width=1.8\columnwidth]{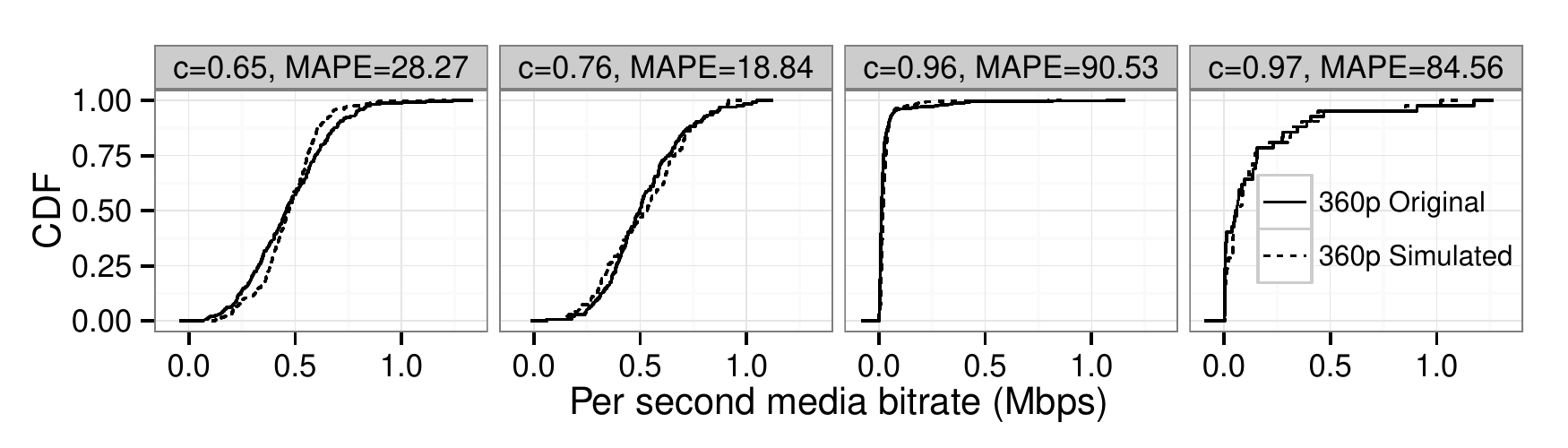}
    } }
    \caption{Results for non-adaptive videos for upscaling 360p and downscaling 720p for varying correlation coefficients. Higher correlations show better fits. High MAPE values show up due to spikes in bit rates, and can be seen at the upper ends of the CDFs.  }
    \label{fig_ecdfunfrag}
  \end{figure*}

Figures~\ref{fig_ecdfhd} and ~\ref{fig_ecdfunfrag} show sample CDF comparisons for DASH HD and non-adaptive case respectively, showing 4 videos for each case. The videos are picked to represent different durations and correlation coefficients (c). Note that we specifically picked some worst case scenarios to demonstrate insights into the behavior, and generally the fits are much better. 

We observed that if c is high, the CDFs fit well even with large MAPE values. The videos we observed for such cases had few high motion peaks. Upscaling resulted in higher peaks than actual and downscaling resulted in lower peaks than actual. The cases with the highest MAPE values in both Figures~\ref{fig_ecdfhd} and ~\ref{fig_ecdfunfrag} represent such videos. We noted that the many of these videos were slide shows of images, commonly seen on YouTube with musical tracks. This is also observed for videos with a stationary background image, because in that case too, there is a spike at segment boundaries to allow seeking and rate shifting. 

\begin{figure*}[!t]
\centerline {
\includegraphics[width=1.8\columnwidth]{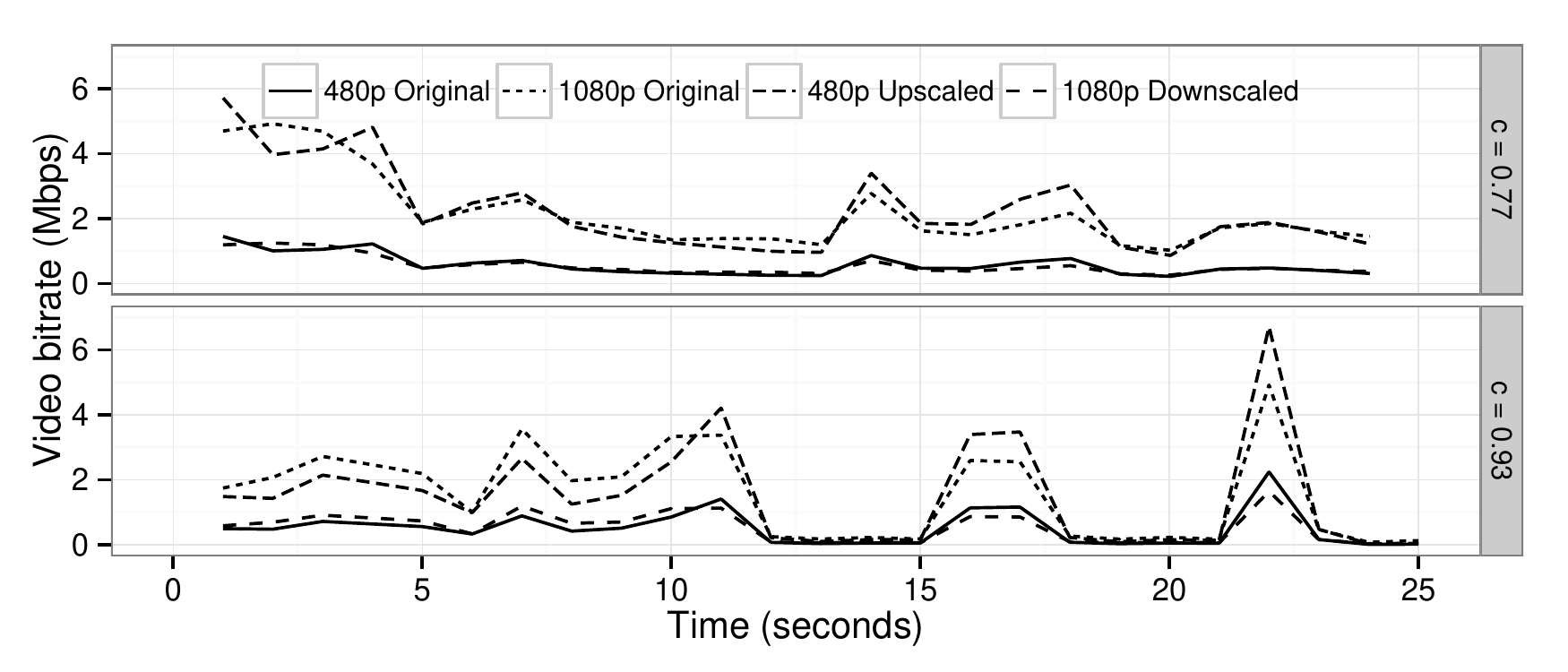}
}
\caption{The video with c=0.77 has MAPE=16.6 for 480p upscaled, and 19.7 for 1080p downscaled. Since this video is not bursty, the out-of-sync peaks go unnoticed in the MAPE values. The video with c=0.93 has MAPE=33.4 for 480p upscaled, and 62.7 for 1080p downscaled. The video has a few widely spaced spikes which are not simulated well, leading to high MAPE. A smoothing function can help overcome this.}
\label{fig_timing}
\end{figure*}

When the correlation coefficients are low, even with lower MAPE values, the CDF fit is not so good. Videos with low c tend to have low MAPE when the video is not bursty, and even though the peaks do not line up, the difference in peaks and troughs is so small that the simulation error is small. Figure \ref{fig_timing} shows the timing graphs of two videos, one with high correlation and high MAPE and the other with low correlation and low MAPE.

\begin{table}[!t]
\centering{
\begin{tabular}{lcc} \hline 
	&\multicolumn{2}{c}{\textbf{MAPE}} \\
	&(mean)& (95\%) \\ \hline
\multicolumn{3}{c}{\textbf{DASH}}\\ \hline
480p to 1080p & 16.51\% & 33.40\% \\
1080p to 480p& 21.03\% & 47.92\% \\ \hline
360p to 720p &17.70 \% & 39.80\% \\
720p to 360p& 21.14\% & 52.82\% \\ \hline
\multicolumn{3}{c}{\textbf{Non-adaptive}}	\\ \hline
360p to 720p & 20.96 \% & 42.87 \% \\
720p to 360p& 31.71\% & 84.98\% \\ \hline
\end{tabular}
}
\caption{The mean and 95th percentile of Mean Absolute percentage Error (MAPE) for the simulations.}
\label{tab_mape}
\end{table}

\section{Discussion}
\label{discussion}
We looked at the characteristics of Internet video in light of large scale active measurements to measure Internet video performance at the endpoint. Since the true user experience should be based on user behavior, active measurements should be based on videos that are popular amongst users. YouTube provides a good use case as it is widely used, accessible without requiring a login and provides localized charts. 

Previous studies show that user experience for Internet video based on TCP streams directly depends on the number and duration of stalls during playout\cite{hossfeld2011quantification,zink2009characteristics, huang12imc,casas2013monitoring,juluri2013viewing,plissonneau2012analyzing}. This is understandable, since deterioration due to packet loss does not occur in this case like it would for say RTP video. So a good approach for measuring performance is to measure stall duration when downloading popular video content and because popularity differs based on geo-location, tests conducted in different locations should use the locally popular videos. This helps in scalability as well, as we do not want millions of active measurement agents flooding measurement servers for a single video. But videos are inherently very different in terms of duration and burstiness, and it is hard to directly compare results based on stalling duration without taking into account these characteristics. A bursty video with high bit rate is more likely to cause stalls than a consistently low bit rate video. Furthermore, an initial buffering of 2 seconds of playout for a 5 second video is likely to prevent stalling when the media bit rate is bordering the capacity of the end-to-end path, while a longer video may still experience stalls. In DASH the player can switch to a higher or lower bit rates representation of the same stream based on the observed path characteristics. This switching can directly affect the user experience if it is done too often and especially if it changes quality by a large magnitude\cite{zink2003subjective}. Hence to measure DASH performance, we need to run tests for a reasonable amount of time and check the stability of the network; here too the burstiness of the video plays a part. 
In the following subsections, we discuss various aspects of our results, their implications and use in the design of active measurements and also their applicability in other fields.

\subsection{Duration of an Active Measurement}

Active measurements must be conducted during idle time, when user traffic is not detected, and must finish as soon as possible so as to avoid affecting any new user-generated traffic. While we want to capture the user experience, we have to do so within these constraints. Video tests, unlike other measurements are bound to run for longer durations depending on the length of the video. We looked at popular videos from 58 different locations and all available categories and found that over 50\% videos have a duration between 1 and 7 minutes. These durations are good enough for active measurements as it gives sufficient time to gauge network performance and are not too long that they start interfering with regular user traffic. Hence, a measurement agent can specifically pick videos that lie within this duration.

A different approach for limiting the duration of the video is to simply cut-off the test at a fixed time. We provide results to prove our hypothesis that testing for the first 3 minutes adequately represents the entire video and so testing can be limited to the first 3 minutes. This also falls in line with the user behavior of aborting videos without watching them to the end~\cite{finamore2011youtube}. However, we must not ignore the fact that these results are based on YouTube videos, and while longer videos are present in the dataset, majority videos are under 10 minutes duration. For movies on Netflix or other Movies on Demand services, the first 3 minutes would be mostly credits having comparatively less motion and hence lower bit rates. Additionally, users are less likely to abort the video in the first few minutes. We will have to run similar tests for such videos to know what their behavior is. Even if the first 3 minutes are not adequate, it is still not feasible to run active tests for entire video durations for long videos. It may be more appropriate, however, to pick a 3 minute or 5 minute segment from the middle of the movie rather than the beginning. 

\subsection{Format of the Video}

Since the same video is available in a number of formats, a measurement agent should be aware of which format to download for performing active measurements. The choice should portray user behavior, and we can relate it to the behavior of players in popular browsers or applications. At the time of writing this paper, WebM is supported in the latest versions of Opera, Google Chrome and Mozilla Firefox, and on Internet Explorer with an additional component~\cite{webm-projsupport}. However, Safari still has no support for WebM. Similarly, MPEG DASH support is also being added to browsers. It is not in the scope of our paper to study the usage trends of these browser or other video applications and to comment on most commonly used formats. However, from our data we observe that both WebM and MP4 files have comparable file sizes and bit rates, but MP4 has slightly smaller media bit rates. The observed difference is small and we expect the results from either format to be comparable to the other because the compression efficiency depends on encoding parameters of the respective codec and the efficiency can be adequately adjusted by appropriately tweaking the codec parameters.

\subsection{Video Category}

We noticed that the popularity of music videos is one that is least affected by geographic locations. While, there are music videos that are popular only in a specific geographic location, a large number of videos were common to charts in various different locations. This also shows up in the incredibly high view counts for a relatively small number of videos collected in our dataset. A measurement scheme that uses music charts only can generate more consistent and comparable results as most music videos have similar lengths and content. 80\% of the music videos in our own dataset had durations between 2 and 6 minutes. This scheme may suffer on grounds of scalability, because a large number of MAs from different locations would be measuring the same video and hence may affect popularity indices. Also, testing a specific category may limit diversity as well, as not all categories of videos watched by users will be covered, and this would make the results biased if a service treats traffic for different categories differently. 

\subsection{Generating Video Traffic}

In some cases, it may be useful to use our own test servers for testing video performance e.g. for troubleshooting service or network issues. While this would not directly reflect upon the performance of a particular service and hence real user experience, it still shows the ISPs internal network performance. If bad performance is measured by a measurement agent, testing with a network local server can help narrow down on the source of the problem. We propose a simple traffic generator for different resolutions using a copy of the video in only one resolution and the average bit rates for the rest. In spite of some shortcomings, this is a good step towards traffic generation and can have applications other than performance measurements.

For example, DASH servers typically require multiple representations of the same video in different resolutions, further depending on the use-case (live streaming or video on demand) it may also serve the same video in different chunk or segment sizes (1s, 2s, 5s, 10s, etc.), this leads to a high number of valid combinations. The streaming server may need to store and handle all the valid combinations of the media stream which creates unnecessary complexity for evaluating congestion control algorithms. In this case, the server can instead store a single file of a particular video stream containing the frame sizes and the timestamps. Eventually up/down scaling the frame sizes to the requested bit rate (as described in section~\ref{analysis2}). Similarly, the media traffic generator can also be  applied for evaluating congestion control algorithm in Web-based Real-time Communication (WebRTC)~\cite{draft.rmcat.evaluate,draft.rmcat.eval.test} by using a single representation of the video as the baseline and changing the frame size based on the bit rate computed by the congestion control algorithm.

\subsection{Feedback Loop}

Internet video is constantly evolving to meet new technological advances and user trends. During the course of our 10 month study we observed several changes on YouTube (removal of FLV for 360p, removal of non-adaptive streams for Full HD, appearance of DASH for Full HD and 4K video). Furthermore, YouTube now offers longer and larger videos than it did 4 years ago. If we use location-based popularity charts for selecting videos when testing with live services, LMAP can automatically keep up with the trends. In addition, we can use a feedback loop from our measurements into our video selection process and traffic generator model. It may not be feasible to gather frame-level information for all measurement agents, as it would require first storing this large amount of data and then sending it to a centralized server for processing. But since the measurement agents in one location will be cycling through the same set of popular videos, we can collect detailed frame-level stats at one central point. Needless to say, without a method that ensures evolution, we cannot keep the testing system relevant for long.

\section{Active Measurements Model}
\label{lmapmodel}

\begin{figure}[!t]
\centerline {
\includegraphics[width=3.2in]{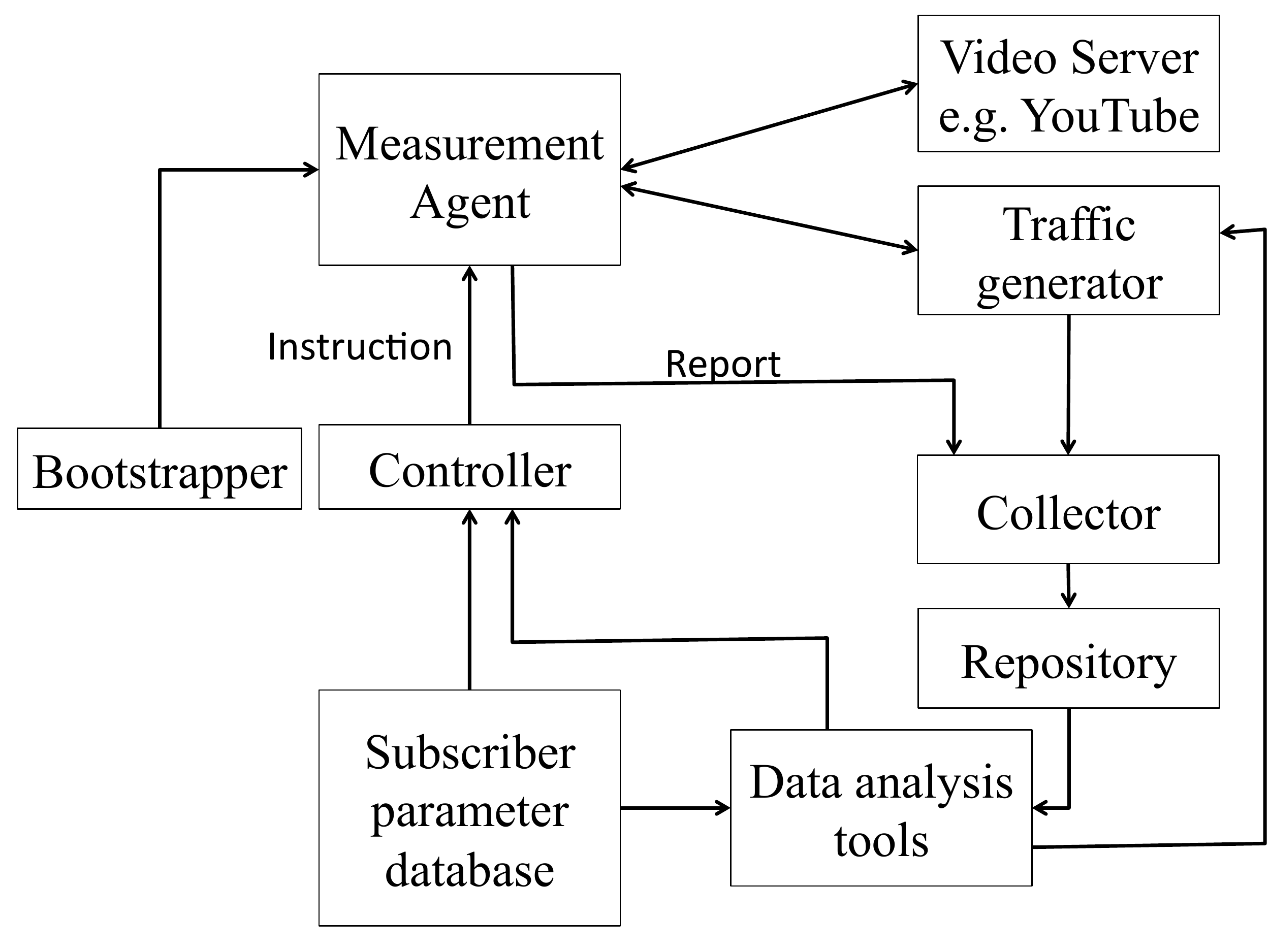}
}
\caption{Main components of an LMAP measurement system for measuring video characteristics.}
\label{fig_lmap}
\end{figure}

The LMAP IETF is working on standardizing large scale performance measurements for access devices. The main components of an LMAP system~\cite{draft-lmap-fw} for active video measurements are shown in Figure \ref{fig_lmap}. The measurement agent (MA) is located at the end-user's premises and is responsible for conducting active tests with YouTube or other video servers. The subscriber database contains information about the subscriber line and the repository is the database of the results. We define the following parameters for our testing model 

\begin{itemize}
  \item M: Number of measurement agents
  \item MINLENGTH : Minimum Length of test videos, value = 72 sec
  \item CUTOFF: Cut-off duration for tests, value = 180 seconds
  \item N : Number of videos to get from charts, value depends on M and frequency of tests \ldots
\end{itemize}

Note that MINLENGTH is the first quartile of the video durations and should be less than CUTOFF. The value of N determines how many videos will eventually be used for testing, and the number should be selected with two considerations 1) It should be much smaller than M so we can get a good number of results from different MAs for the same video. 2) It should not be so small that active measurements start impacting popularity indices of videos. 

There are 3 types of instructions that a controller can request from a MA, which we are already using with our YouTube tests on SamKnows probes: 
 \begin{enumerate}
  \item Collect charts and do random test: The MAs queries top N videos for its location from YouTube, discard videos with duration<MINLENGTH and randomly tests for 1 video. 
  \item Do testing for video URL
\end{enumerate}

The URL can be a YouTube video or it can be a link to the traffic generator, but currently we do not have a traffic generator within our test system, so only YouTube videos are used. In the presence of a traffic generator, the testing methodology consists of the following steps.
\begin{itemize}
  \item Once a week the MAs collect charts and submit results for 1 random video with duration\textgreater MINLENGTH, to the repository via the collector. 
  \item The data analysis tools assign categories to these videos based on duration and bit rates and submits the final list to the controller and the traffic generator.
  \item The traffic generator downloads low-bit rate formats for the selected videos and saves frame logs, it further assigns burstiness values to the videos based on the standard deviation in instantaneous bit rates. It submits burstiness results to the collector which are then incorporated in the categorization done by the data analysis tools. 
  \item  For the remainder of the week the controller randomly assigns videos to different MAs for testing and collect results.
  \item The data analysis tools keep the value of MINLENGTH updated based on duration of videos used for testing. If MINLENGTH becomes larger than CUTOFF, it is ignored while collecting charts. 
  \ldots
\end{itemize}

The traffic generator is used for cases when a controlled video server is needed, for instance, when troubleshooting a problem. It uses frame logs of lowest bit rate streams to generate dummy traffic for testing and uses upscaling for generating traffic for higher resolutions. For upscaling, it needs the average bit rate of high resolution videos. This information can easily be extracted using just the headers from the containers that are present in the beginning of the files, so the whole file is not needed.

\section{Conclusion}
\label{conclusion}

While there are many challenges involved in designing active measurements for
Internet video, the one tricky question that faces measurement designers is
what video to download. We need the measurements at different end points to
somehow correlate with each other, but downloading the same video would defeat
the purpose as it would not reflect true user experience. Downloading popular
(most viewed) videos is the obvious choice for measuring user experience, but
then how do we correlate the performance of a 1 minute video with a 1 hour
video, or a 2Mbps video with an 8Mbps video. For this reason, it is necessary
to know what is a sane value for duration and media bit rate that effectively
represents the majority of the popular videos on the Internet. In light of our
analysis, we conclude that 3 minutes, which is about the median of the video
durations, is a good cut-off duration for our active measurements. As active
measurements are conducted at the end-user, preferably in the absence of
cross-traffic generated from within the user's network, the length of a single
test must be short enough to be conducted conveniently without disrupting the
user. The 3 minute duration covers a wide range of online videos without being
so long that downloading it on a cross-traffic free line becomes a challenge
on its own.

Furthermore, we now know that majority of 1080p videos have bit rates no higher
than 5 to 7Mbps, and though videos with higher bandwidth requirements exist
they are very few in number. Hence, it might be sufficient to measure for
videos that have bit rates in this range. Using these guidelines and picking
videos from a set of current popular videos for the location of the measurement
device, would also make the measurements scalable. Given we pick from a large
enough dataset, the chances of increasing the popularity of a single video just
because we are running active measurements on it is reduced.

The correlation in the file sizes for WebM and MP4 also indicates that the
average media bit rate for both formats is comparable and hence the performance
of either of the two formats should be enough to gauge the user experience. A
next step to consider in this direction would be to correlate media bit rate
variations for the different codecs as well.

Finally, looking at the standard deviation in the instantaneous media bit rates
gives a very simple method for defining burstiness in the video. This paper
just provides a preliminary study of this behavior, but if the standard
deviation can be successfully correlated with performance under restricted
network bandwidth, we can use it to further categorize videos for reconciling
measurement results. Furthermore, such information can help optimize MPEG DASH
clients~\cite{sodagar2011mpeg}, for example, shifting to a higher bit rate for
chunks that have high burstiness only when there is a certain duration of
pre-buffered video.

There are many aspects of the collected data that have not been explored by us.
The file sizes and durations can be used for calculating the average bit rates
for the various formats, which may be used for modeling Internet video. The
data can be used to explore social aspects for instance to see the changes in
popularity (number of views) over the two month period in the common set of
videos.

\section{Acknowledgements}
This work was supported by the European Community’s Seventh Framework Programme (FP7/2007-2013) Grant No. 317647 (Leone).

\bibliographystyle{abbrv}
{\small

\bibliography{biblo}
}

\end{document}